\documentclass[12pt,letterpaper]{article}

\usepackage{mystyle}

\newcounter{small_constant}
\newcommand{\nextsc}{\addtocounter{small_constant}{1}  c_{\arabic{small_constant}} }
\newcommand{\nextscnu}{c_{\arabic{small_constant}}}

\newcounter{small_constant_1}

\newcommand{\etX}{\bar{X}}

\begin{document}

\title{Secure Degrees of Freedom of One-hop Wireless Networks\thanks{This work was supported by NSF Grants CNS 09-64632, CCF 09-64645, CCF 10-18185 and CNS 11-47811.}}

\author{Jianwei Xie \qquad Sennur Ulukus\\
\normalsize Department of Electrical and Computer Engineering\\
\normalsize University of Maryland, College Park, MD 20742 \\
\normalsize {\it xiejw@umd.edu} \qquad {\it ulukus@umd.edu}}

\maketitle

\vspace{-0.5cm}

\begin{abstract}
We study the secure degrees of freedom (d.o.f.) of one-hop wireless networks by considering four fundamental wireless network structures: Gaussian wiretap channel, Gaussian broadcast channel with confidential messages, Gaussian interference channel with confidential messages, and Gaussian multiple access wiretap channel. The secrecy capacity of the canonical Gaussian wiretap channel does not scale with the transmit power, and hence, the secure \dof of the Gaussian wiretap channel with no helpers is zero. It has been known that a strictly positive secure \dof can be obtained in the Gaussian wiretap channel by using a helper which sends structured cooperative signals. We show that the exact secure \dof of the Gaussian wiretap channel with a helper is $\frac{1}{2}$. Our achievable scheme is based on real interference alignment and cooperative jamming, which renders the message signal and the cooperative jamming signal \emph{separable} at the legitimate receiver, but \emph{aligns} them perfectly at the eavesdropper preventing any reliable decoding of the message signal. Our converse is based on two key lemmas. The first lemma quantifies the \emph{secrecy penalty} by showing that the net effect of an eavesdropper on the system is that it eliminates one of the independent channel inputs. The second lemma quantifies the \emph{role of a helper} by developing a direct relationship between the cooperative jamming signal of a helper and the message rate. We extend this result to the case of $M$ helpers, and show that the exact secure \dof in this case is $\frac{M}{M+1}$. We then generalize this approach to more general network structures with \emph{multiple messages.} We show that the sum secure \dof of the Gaussian broadcast channel with confidential messages and $M$ helpers is 1, the sum secure \dof of the two-user interference channel with confidential messages is $\frac{2}{3}$, the sum secure \dof of the two-user interference channel with confidential messages and $M$ helpers is 1, and the sum secure \dof of the $K$-user multiple access wiretap channel is $\frac{K(K-1)}{K(K-1)+1}$.
\end{abstract}

\newpage

\section{Introduction}

We study secure communications in one-hop wireless networks from an information-theoretic point of view. Wyner introduced the wiretap channel \cite{wyner}, in which a legitimate transmitter wishes to send a message to a legitimate receiver secret from the eavesdropper. The capacity-equivocation region was originally found for the degraded wiretap channel by Wyner \cite{wyner}, then generalized to the general wiretap channel by Csiszar and Korner \cite{csiszar}, and extended to the Gaussian wiretap channel by Leung-Yan-Cheong and Hellman \cite{gaussian}. Multi-user versions of the wiretap channel have been studied recently, e.g., broadcast channels with confidential messages \cite{secrecy_ic, xu_bounds_bc_cm_it_09}, multi-receiver wiretap channels \cite{fading1, ersen_eurasip_2009, bagherikaram_bc_2008} (see also a survey on extensions of these to MIMO channels \cite{ersen_jcn_2010}), two-user interference channels with confidential messages \cite{secrecy_ic, he_outerbound_gic_cm_ciss_09}, multiple access wiretap channels \cite{tekin_gmac_w, cooperative_jamming, ersen_mac_allerton_08, liang_mac_cm_08 ,ersen_mac_book_chapter}, relay eavesdropper channels \cite{relay_1, relay_2, relay_3, relay_4, he_untrusted_relay, ersen_crbc_2011}, compound wiretap channels \cite{ compound_wiretap_channel, ersen_ulukus_degraded_compound}. Since in most multi-user scenarios it is difficult to obtain the exact secrecy capacity region, achievable secure degrees of freedom (d.o.f.) at high signal-to-noise ratio (SNR) cases have been studied for several channel structures, such as the $K$-user Gaussian interference channel with confidential messages \cite{koyluoglu_k_user_gic_secrecy, he_k_gic_cm_09}, the $K$-user interference channel with external eavesdroppers \cite{xie_k_user_ia_compound}, the Gaussian wiretap channel with one helper  \cite{secrecy_ia_new,xiang_he_thesis}, the Gaussian multiple access wiretap channel \cite{secrecy_ia5,raef_mac_it_12}, and the wireless $X$ network \cite{secrecy_ia1}.

In the Gaussian wiretap channel, the secrecy capacity is the difference between the channel capacities of the transmitter-receiver and the transmitter-eavesdropper pairs. It is well-known that this difference does not scale with the SNR, and hence the secure \dof of the Gaussian wiretap channel is zero, indicating a severe penalty due to secrecy in this case. Fortunately, this does not hold in multi-user scenarios. In a multi-user network, focusing on a specific transmitter-receiver pair, other (independent) transmitters can be understood as helpers which can improve the individual secrecy rate of this specific pair by cooperatively jamming the eavesdropper \cite{tekin_gmac_w, cooperative_jamming, wiretap_channel_with_one_helper, ersen_mac_book_chapter}.\footnote{Note that, if reliability was the only concern, then in order to maximize the reliable rate of a given transmitter-receiver pair, all other independent transmitters must remain silent. However, when secrecy in addition to reliability is a concern, then independent helpers can improve the secrecy rate of a given transmitter-receiver pair by transmitting signals \cite{tekin_gmac_w, cooperative_jamming, wiretap_channel_with_one_helper, ersen_mac_book_chapter}.} These cooperative jamming signals also limit the decoding performance of the legitimate receiver. It is also known that if the helper nodes transmit independent identically distributed (i.i.d.) Gaussian cooperative jamming signals in a Gaussian wiretap channel, then the secure \dof is still zero \cite{wiretap_channel_with_one_helper, tekin_gmac_w, cooperative_jamming, raef_mac_it_12}. Such i.i.d. Gaussian signals, while maximally jam the eavesdropper, also maximally hurt the legitimate user's decoding capability. Therefore, we expect that strictly positive secure \dof may be achieved with some {\it weak} jamming signals. Confirming this intuition, \cite{secrecy_ia_new,xiang_he_thesis} achieved positive secure \dof by using nested lattice codes in a Gaussian wiretap channel with a helper. In this paper, we obtain the exact secure \dof of several Gaussian network structures, including the Gaussian wiretap channel with a helper, by characterizing this trade-off in the cooperative jamming signals of the helpers.

\begin{figure}[t]
\centering
\includegraphics[scale=0.8]{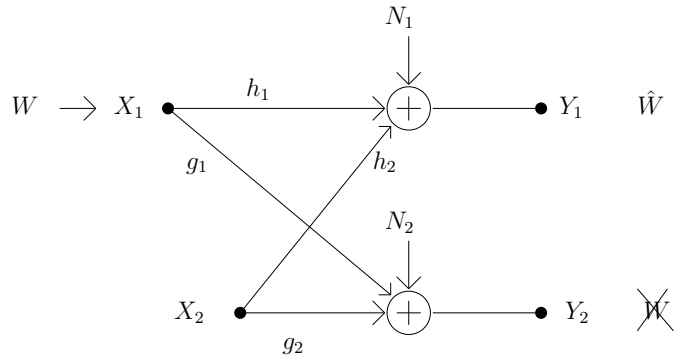}
\caption{Gaussian wiretap channel with one helper.}
\label{fig:gwc_one_helper}
\end{figure}

We start by considering the Gaussian wiretap channel with a single helper, as
shown in Figure~\ref{fig:gwc_one_helper}. In this channel model, secure \dof
with i.i.d. Gaussian cooperative signals is zero
\cite{wiretap_channel_with_one_helper}, and strictly positive secure \dof can be
obtained, for instance, by using nested lattice codes
\cite{secrecy_ia_new,xiang_he_thesis}. Considering this model as a special case
of other channel models, we can verify that $\frac{1}{4}$ secure \dof can be
achieved as a symmetric individual rate on the two-user interference channel
with external eavesdroppers \cite{xie_k_user_ia_compound} and on the multiple
access wiretap channel \cite{secrecy_ia5}. References \cite{secrecy_ia4} and
\cite[Theorem 5.4 on page 126]{xiang_he_thesis} showed that with integer lattice
codes a secure \dof of $\frac{1}{2}$ can be achieved if the channel gains are
\emph{irrational algebraic numbers.} While such class of channel gains has zero
Lebesgue measure, the idea behind this achievable scheme can be generalized to
much larger set of channel gains. The enabling idea behind this achievable scheme is
as follows: If the cooperative jamming signal from the helper and the message signal from the
legitimate user can be aligned in the same \emph{dimension} at the eavesdropper,
then the secrecy penalty due to the information leakage to the eavesdropper can
be upper bounded by a constant, while the information transmission rate to the
legitimate user can be made to scale with the transmit power. Following this
insight, we propose an achievable scheme based on real interference alignment
\cite{real_inter_align, real_inter_align_exploit} and cooperative jamming to
achieve $\frac{1}{2}$ secure \dof for \emph{almost all channel gains.} This constitutes
the best known achievable secure \dof for the Gaussian wiretap channel with a
helper. The cooperative jamming signal from the helper can be distinguished from
the message signal at the legitimate receiver by properly designing the
structure of the signals from both transmitters; meanwhile, they can be aligned
together at the observation space of the eavesdropper to ensure
undecodability of the message signal, hence secrecy (see Figure~\ref{fig:gwc_one_helper_ia}).
Intuitively, the end result of $\frac{1}{2}$ secure \dof comes from the facts that the cooperative
jamming signal and the message signal should be of about the same size to align at the eavesdropper,
and they should be separable at the legitimate receiver, who can decode at most a total of $1$ d.o.f.
We analyze the rate and equivocation achieved by this scheme by using the Khintchine-Groshev theorem
of Diophantine approximation in number theory.

For the converse for this channel model, the best known upper bound is
$\frac{2}{3}$ \cite[Theorem 5.3 on page 126]{xiang_he_thesis} which was obtained
by adding virtual nodes to the system and using the upper bound developed in
\cite{interference_alignment_compound_channel}. Reference
\cite{interference_alignment_compound_channel} developed upper bounds for the
secure \dof of the multiple-antenna compound wiretap channel by exploring the
correlation between the $n$-letter observations of a group of legitimate
receivers and a group of eavesdroppers, instead of working with single-letter
expressions. Our converse works with $n$-letter observations as well. Our
converse has two key steps. First, we upper bound the secrecy rate by the
difference of the sum of differential entropies of the channel inputs of the
legitimate receiver and the helper and the
differential entropy of the eavesdropper's observation. This shows that, the
secrecy penalty due to the eavesdropper's observation is tantamount to
eliminating one of the independent channel inputs. As a result, the final upper
bound involves only the  differential entropy of the channel input of the
independent helper. In the second step, we develop a relationship between the
cooperative jamming signal from the independent helper and the message rate. The
goal of the cooperative jamming signal is to further confuse the eavesdropper.
However, the cooperative jamming signal appears in the channel output of the
legitimate user also. Intuitively, if the legitimate user is to reliably decode
the message signal which is mixed with the cooperative jamming signal, there
must exist a constraint on the cooperative jamming signal. Our second step
identifies this constraint by developing an upper bound on the differential
entropy of the cooperative jamming signal in terms of the message rate. These
two steps give us an upper bound of $\frac{1}{2}$ secure \dof for the Gaussian
wiretap channel with a helper, which matches our achievable lower bound. This
concludes that the exact secure \dof of the Gaussian wiretap channel with a
helper is $\frac{1}{2}$ for \emph{almost all channel gains}.

We then generalize our result to the case of $M$ independent helpers. We show
that the exact secure \dof in this case is $\frac{M}{M+1}$. Our achievability extends our
original achievability for the one-helper case in the following manner:
The transmitter sends its message by employing $M$ independent sub-messages, and
the $M$ helpers send independent cooperative jamming signals. Each cooperative
jamming signal is aligned with one of the $M$ sub-messages at the eavesdropper to
ensure secrecy (see Figure~\ref{fig:gwc_m_helper_ia}). Therefore, each
sub-message is protected by one of the $M$ helpers. Our converse is an
extension of the converse in the one-helper case. In particular, we upper bound the secrecy
rate by the difference of the sum of the differential entropies of all of the channel
inputs and the differential entropy of the eavesdropper's observation. The secrecy penalty
due to the eavesdropper's observation eliminates one of the channel inputs, which we choose as the legitimate user's channel input. We then utilize the relationship we developed between the differential entropy of each of the cooperative jamming signals and the message rate. The upper bound so developed matches the achievability lower bound, giving the exact secure \dof for
the $M$-helper case.

As an important extension of the single-message one-helper problem, we
consider the broadcast channel with confidential messages and one-helper, where a transmitter
wishes to send two messages securely to two users on a broadcast channel while keeping each message
secure from the unintended receiver. Without a helper, the sum secure \dof of this channel model is zero.
We show that with one helper, the exact sum secure \dof is $1$. The sum secure \dof remains the same
as more helpers are added. The achievability for the one-helper case is as follows: The transmitter sends the channel input by putting two messages on different \emph{rational dimensions}. Meanwhile, the cooperative jamming signal from the helper is designed in such a way that it aligns with the unintended message, but leaves the intended message intact, at each receiver (see Figure~\ref{fig:gbc_one_helper_ia}). The converse for this case
follows from the converse without any secrecy constraints for the Gaussian broadcast channel, which is $1$.

Cooperative jamming based achievable schemes are intuitive for the
independent-helper problems due to the fact that the helpers do not have
messages of their own. Such schemes can be extended to multiple-transmitter (with independent messages)
settings, such as, interference channels with confidential messages and multiple
access wiretap channel, etc. All previous works extended this approach in the
following way: Each transmitter simply sends one message signal, and the message
signals from all of the transmitters are \emph{aligned} together at the
eavesdropper. Due to the mixture of the message signals, the eavesdropper is
confused regarding any one of the message signals, and a positive secure \dof is
achievable. However, this approach is sub-optimal. To achieve optimal secure
d.o.f., we need to design the structure of the channel inputs more carefully.
We propose the following transmission structure: Besides the message carrying
signal, each transmitter also sends a cooperative jamming signal. The exact
number and the structure of the message signals and the cooperative jamming signals
depend on the specific network structure.

For the two-user Gaussian interference channel with confidential messages,
previously known lower bounds for the sum secure \dof are $\frac{1}{3}$
\cite{secrecy_ia1} and $0$ \cite{koyluoglu_k_user_gic_secrecy}, which come from
the general results for the $K$-user case: 
$\frac{K-1}{2K-1}$ \cite{secrecy_ia1} and $\frac{K-2}{2K-2}$
\cite{koyluoglu_k_user_gic_secrecy}. The individual secure \dof
of $\frac{1}{2}$ achieved in \cite{secrecy_ia4} and \cite[Theorem 5.4 on page
126]{xiang_he_thesis} in the context of the wiretap channel with a helper (for
the class of algebraic irrational channel gains) can also be understood as a
lower bound for the sum secure \dof for the two-user interference channel with
confidential messages. We show that, by using interference alignment and
cooperative jamming at both transmitters, we can achieve a sum secure \dof of
$\frac{2}{3}$ for \emph{almost all channel gains}, which is better than all previously
known achievable secure \dof We design an achievable scheme in which each
transmitter sends a mixed signal containing the message signal and a cooperative
jamming signal. These two components have the same signaling structure, and are
separable at the intended receiver. Furthermore, the cooperative jamming signal
is perfectly \emph{aligned} with the message signal from the other transmitter
(see Figure~\ref{fig:gic_cm_ia}).  Our converse
starts with considering transmitter $2$ as a helper for transmitter-receiver
pair $1$. In contrast to the single-message case, since transmitter $2$ also
intends to deliver a message $W_2$ to receiver $2$, in the second step, we treat
transmitter $1$ as the helper for the transmitter-receiver pair $2$ and upper
bound the differential entropy of its channel input by using its relationship
with the message rate of $W_2$. The converse matches the achievability lower
bound, giving the exact secure \dof for the two-user interference channel with
confidential messages as $\frac{2}{3}$.

We then generalize this result to the case with one helper, i.e., two two-user
Gaussian interference channel with confidential messages and one helper. We show
that a sum secure \dof of $1$ is achievable. The structure of the channel inputs
in the corresponding achievable scheme is simpler than in the cases of previous
channel models. Each transmitter sends a signal carrying its message. With
probability one, these two signals are not in the same \emph{rational dimension}
at the receivers. On the other hand, the cooperative jamming signal from the
helper can be aligned with the unintended message at each receiver while leaving
the intended message intact (see Figure~\ref{fig:gic_cm_m_helper_ia}). The
converse for this case follows from the converse without any secrecy constraints
for the two-user Gaussian interference channel
\cite{multiplexing_gain_of_networks}, which is $1$.  This concludes that the
exact sum secure \dof of the two-user Gaussian interference channel with
confidential messages and one helper is $1$. Since utilizing one helper is
sufficient to achieve the upper bound, the sum secure \dof remains the same for
arbitrary $M$ helpers.

For the $K$-user multiple access wiretap channel, the best known lower bound for the sum secure \dof is
$\frac{K-1}{K}$ \cite{secrecy_ia5} which gives $\frac{1}{2}$ for $K=2$. In
addition, for $K=2$, the individual secure \dof of $\frac{1}{2}$ achieved in
\cite{secrecy_ia4} and \cite[Theorem 5.4 on page 126]{xiang_he_thesis} in the
context of the wiretap channel with a helper (for the class of algebraic
irrational channel gains) can also be understood as a lower bound for the sum
secure \dof for the two-user multiple access wiretap channel.
We show that, by using interference alignment and cooperative jamming at all
transmitters simultaneously, we can achieve a sum secure \dof of
$\frac{K(K-1)}{K(K-1)+1}$ for the $K$-user multiple access wiretap channel, for
\emph{almost all channel gains}, which is better than all previously known achievable
secure \dof In particular, for $K=2$, our achievable scheme gives a sum secure
\dof of $\frac{2}{3}$. In order to obtain this sum secure d.o.f., we need a more
detailed structure for each channel input. Each transmitter sends a mixed signal
containing the message
signal and a cooperative jamming signal. Specifically, each transmitter divides its own
message into $K-1$ sub-messages each of which having the same structure as the
cooperative jamming signal. By such a scheme, the total $K$ cooperative jamming
signals from the $K$ transmitters \emph{span} the whole \emph{space} at the
eavesdropper's observation, in order to hide each one of the message signals from the
eavesdropper. On the other hand, to maximize the sum secrecy d.o.f., the
cooperative jamming signals from all of the transmitters are \emph{aligned} in
the same \emph{dimension} at the legitimate receiver to occupy the smallest
\emph{space} (see Figure~\ref{fig:mac_k_ia}). Our converse is a generalization of
our converse used in earlier channel model. We first show that the sum secrecy rate is upper
bounded by the sum of differential entropies of all channel inputs except the
one eliminated by the eavesdropper's observation. Then, we consider each channel input
as the jamming signal for all other transmitters and upper bound
its differential entropy by using its relationship with the sum rate of the messages
belonging to all other transmitters. This gives us a matching converse and shows
that the exact sum secure \dof for this channel model is $\frac{K(K-1)}{K(K-1)+1}$.

\section{System Model and Definitions}
In this paper, we consider four fundamental channel models: wiretap channel with
helpers, broadcast channel with confidential messages and helpers, two-user
interference channel with confidential messages and helpers, and multiple access
wiretap channel. In this section, we give the channel models and relevant
definitions. All the channels are additive white Gaussian noise (AWGN) channels.
All the channel gains are time-invariant, and independently drawn from
continuous distributions.

\subsection{Wiretap Channel with Helpers}
\label{sec:gwch_model_gwch}

\begin{figure}[t]
\centering
\includegraphics[scale=0.8]{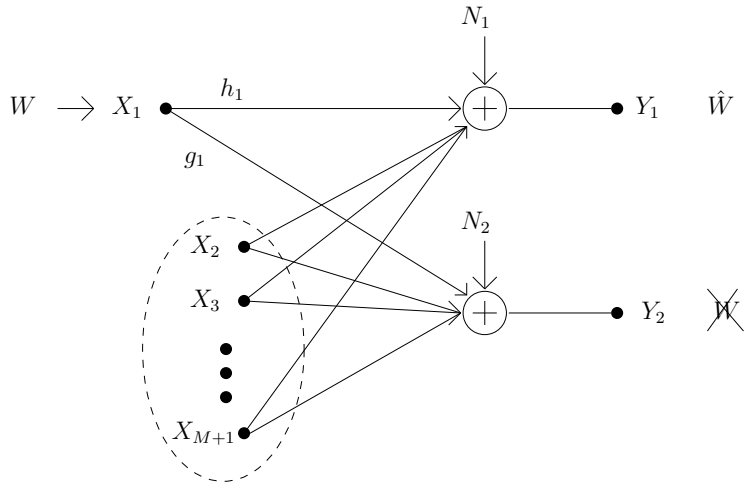}
\caption{Gaussian wiretap channel with $M$ helpers.}
\label{fig:gwc_helper_general}
\end{figure}

The Gaussian wiretap channel with helpers (see Figure~\ref{fig:gwc_helper_general}) is defined by,
\begin{align}
\label{eqn:gwch_channel_model_helpers_genneral_1}
Y_1 &= h_1 X_1 + \sum_{j=2}^{M+1}h_j X_j + N_1 \\
\label{eqn:gwch_channel_model_helpers_genneral_2}
Y_2 & = g_1 X_1 + \sum_{j=2}^{M+1}g_j X_j + N_2
\end{align}
where $Y_1$ is the channel output of the legitimate receiver, $Y_2$ is the
channel output of the eavesdropper, $X_1$ is the channel input of the legitimate
transmitter, $X_i$, for $i=2,\ldots,M+1$, are the channel inputs of the $M$
helpers, $h_i$ is the channel gain of the $i$th transmitter to the legitimate
receiver, $g_i$ is the channel gain of the $i$th transmitter to the
eavesdropper, and $N_1$ and $N_2$ are two independent zero-mean unit-variance
Gaussian random variables. All channel inputs satisfy average power constraints,
$\E\left[X^2_{i}\right] \le P$, for $i=1,\ldots, M+1$. 

Transmitter $1$ intends to send a message $W$, uniformly chosen from a set
$\mathcal{W}$, to the legitimate receiver (receiver $1$). The rate of the
message is $R\defn\frac{1}{n}\log|\mathcal{W}|$, where $n$ is the number of
channel uses. Transmitter $1$ uses a stochastic function $f: \mathcal{W}\to
\bfX_1$ to encode the message, where $\bfX_1\defn X_1^n$ is the $n$-length
channel input.\footnote{We use boldface letters to denote $n$-length vector
signals, e.g., $\bfX_1\defn X_1^n$, $\bfY_1\defn Y_1^n$, $\bfY_2\defn Y_2^n$, etc.} The legitimate
receiver decodes the message as $\hat{W}$ based on its observation $\mathbf{Y}_1
$. A secrecy rate $R$ is said to be achievable if for any $\epsilon>0$ there
exists an $n$-length code such that receiver $1$ can decode this message
reliably, i.e., the probability of decoding error is less than $\epsilon$,
\begin{equation}
\label{eqn:reliability_measure}
\pr\left[W\neq\hat{W}\right] \le \epsilon
\end{equation}
and the message is kept information-theoretically secure against the eavesdropper,
\begin{equation}
\label{eqn:secrecy_measure}
\frac{1}{n}H(W| \bfY_2) \ge \frac{1}{n} H(W) - \epsilon
\end{equation}
i.e., that the uncertainty of the message $W$, given the observation $\bfY_2$ of
the eavesdropper, is almost equal to the entropy of the message. The supremum of
all achievable secrecy rates is the secrecy capacity $C_s$ and the secure
d.o.f., $D_s$, is defined as
\begin{equation}
D_s \defn \lim_{P\to\infty}  \frac{C_s}{\frac{1}{2}\log P}
\label{eqn:sec-dof-defn}
\end{equation}

Note that $D_s\le 1$ is an upper bound. To avoid trivial cases, we assume that $h_1\neq 0$ and $g_1\neq 0$. Without the independent helpers, i.e., $M=0$, the secrecy capacity of the Gaussian wiretap channel is known \cite{gaussian}
\begin{equation}
C_s = \frac{1}{2}\log\left(1+h_1^2 P\right) -  \frac{1}{2}\log\left(1+g_1^2 P\right)
\end{equation}
and from \eqn{eqn:sec-dof-defn} the secure d.o.f.~is zero. Therefore, we assume
$M\ge 1$. If there exists a $j$ ($j=2,\ldots,M+1$) such that $h_j=0$ and
$g_j\neq 0$, then a lower bound of $1$ secure \dof  can be
obtained for this channel by letting this helper jam
the eavesdropper by i.i.d.~Gaussian noise of power $P$ and  keeping all other
helpers silent. This lower bound matches the upper bound, giving the secure \dof
On the other hand, if
there exists a $j$ ($j=2,\ldots,M+1$) such that $h_j\neq0$ and $g_j= 0$, then
this helper can be removed from the channel model without affecting the secure
d.o.f. Therefore, in the rest of the paper, for the case of Gaussian wiretap
channel with $M$ helpers, we  assume that $M\ge1$ and $h_j\neq0$ and $g_j\neq 0$
for all $j=1,\cdots,M+1$.

\subsection{Broadcast Channel with Confidential Messages and Helpers}
\label{sec:gwch_model_bch}

\begin{figure}[t]
\centering
\includegraphics[scale=0.8]{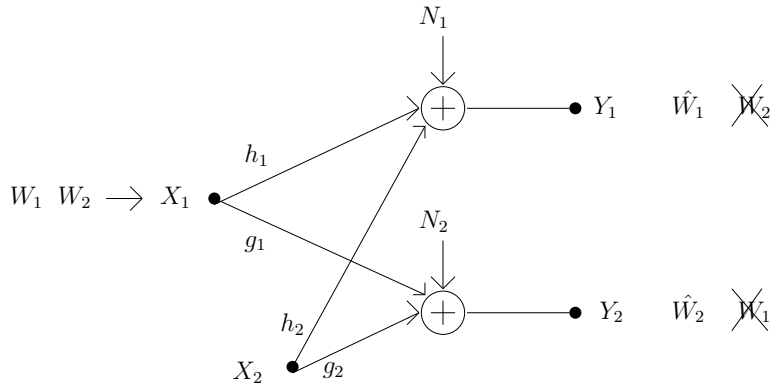}
\caption{Gaussian broadcast channel with confidential messages and $M=1$ helper.}
\label{fig:gbc_one_helper}
\end{figure}

The Gaussian broadcast channel with confidential messages and helpers (see Figure~\ref{fig:gbc_one_helper} for one helper) is defined by,
\begin{align}
Y_1 &= h_1 X_1 + \sum_{j=2}^{M+1}h_j X_j + N_1 \\
Y_2 & = g_1 X_1 + \sum_{j=2}^{M+1}g_j X_j + N_2
\end{align}
In this model, transmitter $1$ has two independent
messages, $W_1$ and $W_2$, intended for receivers $1$ and $2$, respectively.
Messages $W_1$ and $W_2$ are independently and uniformly chosen from sets $\mathcal{W}_1$ and $\mathcal{W}_2$, respectively. The rates of the messages are $R_1\defn\frac{1}{n}\log|\mathcal{W}_1|$ and $R_2\defn\frac{1}{n}\log|\mathcal{W}_2|$. Transmitter $1$ uses a stochastic function $f: \mathcal{W}_1 \times \mathcal{W}_2 \to \bfX_1$ to encode the messages. The messages are said to be confidential if only the intended receiver can decode each message, i.e., each receiver is an eavesdropper for the other. Transmitters
$2,3,\cdots,M+1$ are the independent helpers. Similar to \eqn{eqn:reliability_measure} and \eqn{eqn:secrecy_measure}, we define the reliability and secrecy of the messages as,
\begin{align}
\pr[W_1\neq\hat{W}_1] & \le \epsilon \\
\pr[W_2\neq\hat{W}_2] & \le \epsilon \\
\frac{1}{n}H(W_1| \bfY_2) & \ge \frac{1}{n} H(W_1) - \epsilon \\
\frac{1}{n}H(W_2| \bfY_1) & \ge \frac{1}{n} H(W_2) - \epsilon
\end{align}
The sum secure \dof for this channel model is defined as
\begin{equation}
D_{s,\Sigma} \defn \lim_{P\to\infty} \sup \frac{R_{1}+ R_{2}}{\frac{1}{2} \log P}
\end{equation}
where the supremum is over all achievable secrecy rate pairs $(R_{1}, R_{2})$.

\subsection{Interference Channel with Confidential Messages and Helpers}
\label{sec:gwch_model_ich}

\begin{figure}[t]
\centering
\includegraphics[scale=0.7]{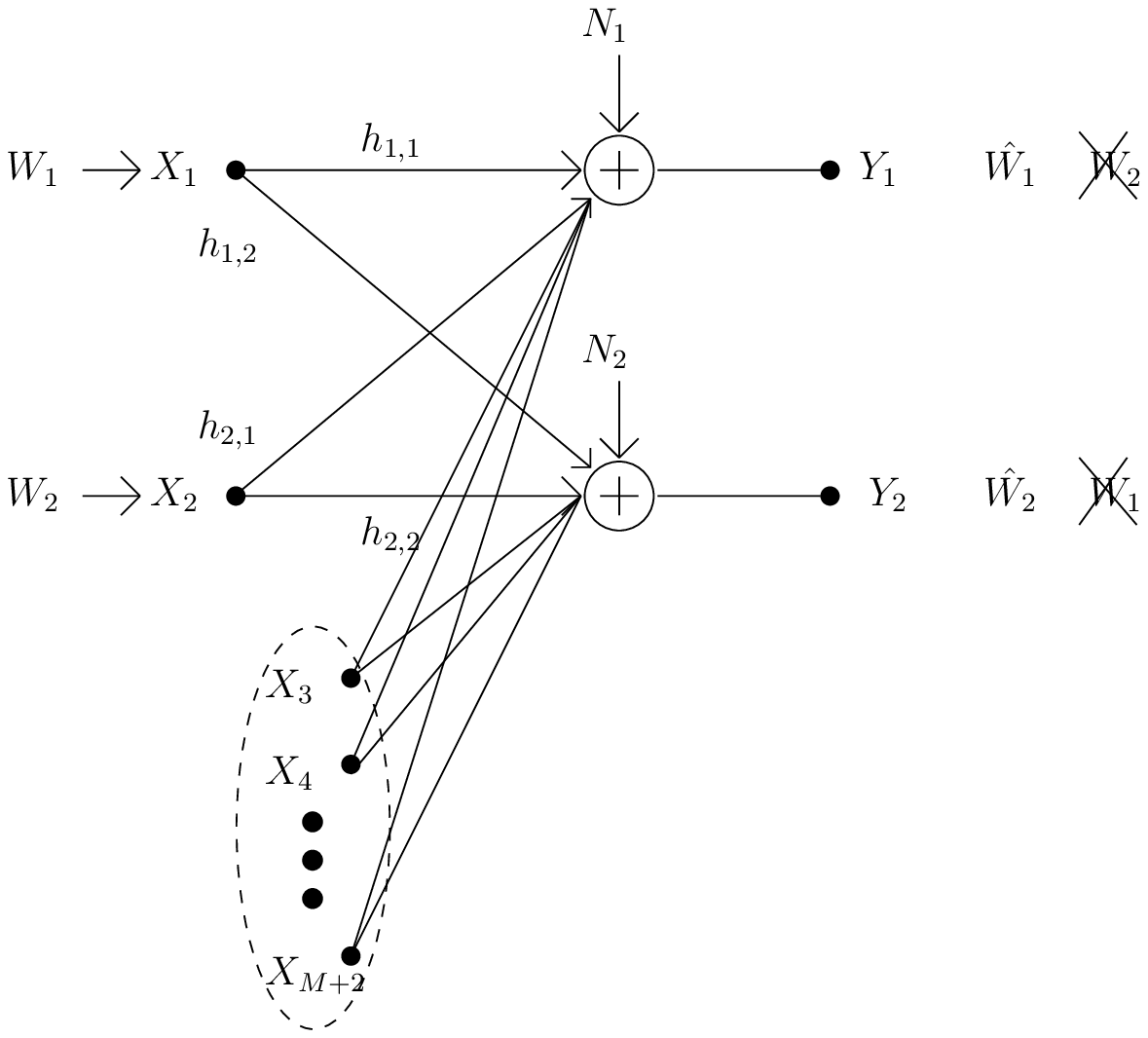}
\caption{Two-user Gaussian interference channel with confidential messages and
$M$ helpers.}
\label{fig:gic_cm_m_helper}
\end{figure}

The two-user Gaussian interference channel with confidential messages and helpers (see Figure~\ref{fig:gic_cm_m_helper}) is defined by,
\begin{align}
Y_1 & = h_{1,1} X_1 + h_{2,1} X_2 + \sum_{j=3}^{M+2} h_{j,1} X_j + N_1 \\
Y_2 & = h_{1,2} X_1 + h_{2,2} X_2 + \sum_{j=3}^{M+2} h_{j,2} X_j + N_2
\end{align}
where $X_1, X_2, \cdots, X_{M+2}, N_1$ and $N_2$ are mutually independent.

\begin{figure}[t]
\centering
\includegraphics[scale=0.8]{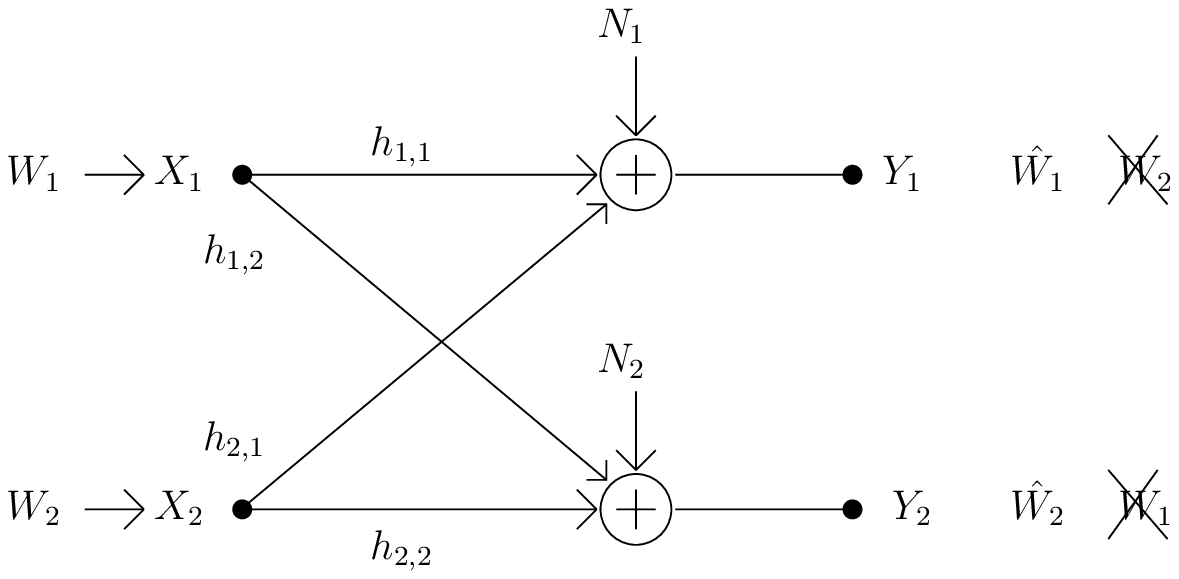}
\caption{Two-user Gaussian interference channel with confidential messages.}
\label{fig:gic_cm}
\end{figure}

One special, but important, case is the two-user Gaussian interference channel with confidential messages, i.e., $M=0$, which is shown in Figure~\ref{fig:gic_cm} and defined  by,
\begin{align}
Y_1 & = h_{1,1} X_1 + h_{2,1} X_2 + N_1 \\
Y_2 & = h_{1,2} X_1 + h_{2,2} X_2 + N_2
\end{align}

In the two-user interference channel with confidential messages, each transmitter wishes to send a confidential message to its own receiver. Transmitter $1$ has
message $W_1$ uniformly chosen from set $\mathcal{W}_1$. The rate of the message is $R_1\defn\frac{1}{n}\log|\mathcal{W}_1|$. Transmitter $1$ uses a stochastic function $f_1: \mathcal{W}_1 \to \bfX_1$ to encode the message. Similarly, transmitter $2$ has
message $W_2$ (independent of $W_1$) uniformly chosen from set $\mathcal{W}_2$. The rate of the message is $R_2\defn\frac{1}{n}\log|\mathcal{W}_2|$. Transmitter $2$ uses a stochastic function $f_2: \mathcal{W}_2 \to \bfX_2$ to encode the message. The messages are said to be confidential if only the intended receiver can decode each message, i.e., each receiver is an eavesdropper for the other. Transmitters
$2,3,\cdots,M+1$ are the independent helpers. Similar to \eqn{eqn:reliability_measure} and \eqn{eqn:secrecy_measure}, we define the reliability and secrecy of the messages as,
\begin{align}
\pr[W_1\neq\hat{W}_1] & \le \epsilon \\
\pr[W_2\neq\hat{W}_2] & \le \epsilon \\
\frac{1}{n}H(W_1| \bfY_2) & \ge \frac{1}{n} H(W_1) - \epsilon \\
\frac{1}{n}H(W_2| \bfY_1) & \ge \frac{1}{n} H(W_2) - \epsilon
\end{align}
The sum secure \dof for this channel model is defined as
\begin{equation}
D_{s,\Sigma} \defn \lim_{P\to\infty} \sup \frac{R_{1}+ R_{2}}{\frac{1}{2} \log P}
\end{equation}
where the supremum is over all achievable secrecy rate pairs $(R_{1}, R_{2})$.

\subsection{Multiple Access Wiretap Channel}
\label{sec:gwch_model_macw}

\begin{figure}[t]
\centering
\includegraphics[scale=0.70]{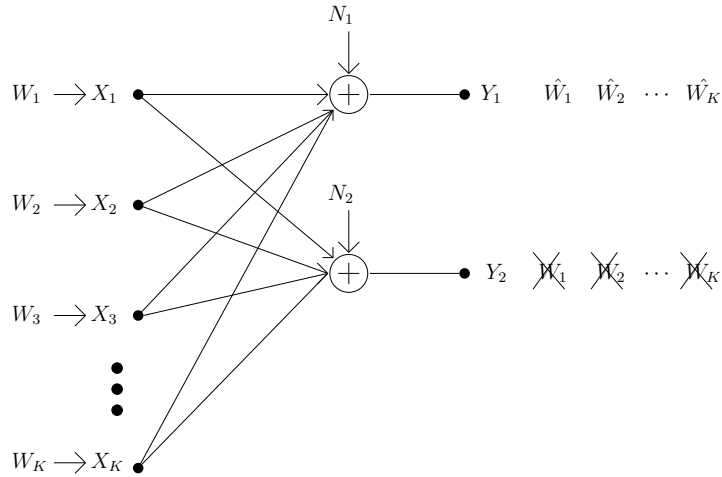}
\caption{$K$-user multiple access wiretap channel.}
\label{fig:mac_k}
\end{figure}

The $K$-user Gaussian multiple access wiretap channel (see Figure~\ref{fig:mac_k}) is defined by,
\begin{align}
\displaystyle Y_1 & = \sum_{i=1}^{K} h_j X_j + N_1 \\
\displaystyle Y_2 & = \sum_{i=1}^{K} g_j X_j + N_2
\end{align}
In this channel model, each transmitter $i$ has a message $W_i$ intended for the
legitimate receiver whose channel output is $Y_1$. All of the messages are
independent. Message $W_i$ is uniformly chosen from set $\mathcal{W}_i$. The
rate of message $i$ is $R_i\defn\frac{1}{n}\log|\mathcal{W}_i|$. Transmitter $i$
uses a stochastic function $f_i: \mathcal{W}_i \to \bfX_i$ to encode its
message.  All of the messages are needed to be kept secret from the eavesdropper, whose channel output is $Y_2$.

Similar to \eqn{eqn:reliability_measure}, the reliability of the messages is defined by
\begin{equation}
\pr\left[(W_1,\cdots, W_K)\neq(\hat{W}_1,\cdots,\hat{W}_K)\right] \le \epsilon
\end{equation}
and similar to \eqn{eqn:secrecy_measure} the secrecy constraint (for the entire message set) is defined as
\begin{equation}
\frac{1}{n} H(W_1, W_2, \cdots, W_K|\bfY_2) \ge
\frac{1}{n} H(W_1, W_2, \cdots, W_K) - \epsilon
\end{equation}
Note that this definition implies the secrecy for any subset of the messages, including individual messages, i.e.,
\begin{align}
 \frac{1}{n} H(W_\mathbf{S}|\bfY_2)
& =  \frac{1}{n} H(W_1, W_2, \cdots, W_K|\bfY_2)
   - \frac{1}{n} H(W_{\mathbf{S}^c}|\bfY_2,W_{{\mathbf{S}}})\\
&  \ge \frac{1}{n} H(W_1, W_2, \cdots, W_K|\bfY_2)
   - \frac{1}{n} H(W_{\mathbf{S}^c}|W_{{\mathbf{S}}})\\
&  \ge \frac{1}{n} H(W_1, W_2, \cdots, W_K) - \epsilon
   - \frac{1}{n} H(W_{\mathbf{S}^c}|W_{{\mathbf{S}}})\\
&  \ge \frac{1}{n} H(W_\mathbf{S}) - \epsilon
\end{align}
for any $\mathbf{S} \subset \{1,2,\cdots,K\}$. The sum secure \dof for this channel model is defined as
\begin{equation}
D_{s,\Sigma} \defn \lim_{P\to\infty} \sup \frac{\sum_{i=1}^{K} R_i}{\frac{1}{2} \log P}
\end{equation}
where the supremum is over all achievable secrecy rate tuples $(R_{1}, \cdots, R_{K})$.

\section{General Converse Results}

In this section, we give two lemmas that will be used in the converse proofs in later
sections.

\subsection{Secrecy Penalty}
\label{sec:gwch_general_ub}
Consider the channel model formulated in Section~\ref{sec:gwch_model_gwch},
where transmitter 1 wishes to have secure communication with receiver 1, in the
presence of an eavesdropper (receiver 2) and $M$ helpers (transmitters 2 through
$M+1$). We propose a general upper bound for the secrecy rate between
transmitter 1 and receiver 1 by working with $n$-letter signals, and introducing
new mutually independent Gaussian random variables $\{\tilde N_i\}_{i=2}^{M}$
which are zero-mean and of variance $\tilde\sigma_{i}^2$ where
$\tilde\sigma_{i}^2<\min(1/h_i^2,1/g_i^2)$, and are independent of all other
random variables. Each vector $\tilde\bfN_i$ is an i.i.d.~sequence of $\tilde
N_i$.

In the following lemma, we give a general upper bound for the secrecy rate.
This lemma states that the secrecy rate of the legitimate pair is upper
bounded by the difference of the sum of differential entropies of all channel
inputs (perturbed by small noise) and the differential entropy of the eavesdropper's observation; see
\eqn{eqn:gwch_general_ub_for_m_helpers}. This upper bound can further be interpreted as follows: If
we consider the eavesdropper's observation as the \emph{secrecy penalty,} then the secrecy penalty is
tantamount to the elimination of one of the channel inputs in the system; see \eqn{eqn:gwch_general_ub_for_m_helpers_kill_y2}.
\begin{lemma}
\label{lemma:gwch_general_ub_for_m_helpers}
The secrecy rate of the legitimate pair is upper bounded as
\begin{align}
n R
& \le \sum_{i=1}^{M+1} h( \tilde\bfX_i)  - h(\bfY_2) +  n c \label{eqn:gwch_general_ub_for_m_helpers} \\
& \le \sum_{i=1,i\neq j}^{M+1} h( \tilde\bfX_i) +  n c' \label{eqn:gwch_general_ub_for_m_helpers_kill_y2}
\end{align}
where  $\tilde\bfX_i = \bfX_i+\tilde\bfN_i$ for $i=1,2,\cdots,M+1$, and $\tilde\bfN_i$ is an i.i.d.~sequence
(in time) of random variables $\tilde N_i$ which are independent Gaussian random variables with zero-mean and variance  $\tilde\sigma_{i}^2$ with $\tilde\sigma_{i}^2<\min(1/h_i^2,1/g_i^2)$. In addition, $c$ and
$c'$ are constants which do not depend on $P$, and $j \in \{1,2,\cdots,M+1\}$
could be arbitrary.
\end{lemma}

\begin{Proof}
We use notation $c_i$, for $i\ge 1$, to denote constants which are independent of the power $P$. We start as follows:
\begin{align}
n R & = H(W) = H(W|\bfY_1) + I(W;\bfY_1) \label{eqn:gwc_converse_start} \\
& \le I(W;\bfY_1)  + n \nextsc \setcounter{small_constant_1}{\arabic{small_constant}}\\
& \le I(W;\bfY_1)  - I(W;\bfY_2) + n \nextsc
\end{align}
where we used Fano's inequality and the secrecy constraint in \eqn{eqn:secrecy_measure}. By providing $\bfY_2$ to receiver $1$, we further upper bound $nR$ as
\begin{align}
n R
& \le I(W;\bfY_1,\bfY_2)  - I(W;\bfY_2) + n \nextscnu \\
& = I(W;\bfY_1|\bfY_2)  + n \nextscnu \\
& = h(\bfY_1|\bfY_2) - h(\bfY_1|\bfY_2,W)  + n \nextscnu \\
& \le h(\bfY_1|\bfY_2)  + n \nextsc \label{eqn:gwc_converse_continue}
\end{align}
where \eqn{eqn:gwc_converse_continue} is due to
\begin{align}
h(\bfY_1|\bfY_2,W) & \ge h(\bfY_1|\bfX_1, \bfX_2, \cdots, \bfX_{M+1}, \bfY_2,W) \\
&=  h(\bfN_1|\bfX_1, \bfX_2, \cdots, \bfX_{M+1}, \bfY_2,W) \\
&=  h(\bfN_1) \\
&= \frac{n}{2}\log 2 \pi e
\end{align}
which is independent of $P$.

In the next step, we introduce random variables $\tilde\bfX_i$ which are noisy versions of the channel inputs $\tilde\bfX_i = \bfX_i+\tilde\bfN_i$ for $i=1,2,\cdots,M+1$. Thus, starting from \eqn{eqn:gwc_converse_continue},
\begin{align}
n R
& \le h(\bfY_1|\bfY_2)  + n \nextscnu \\
& = h(\bfY_1,\bfY_2) - h(\bfY_2)  + n \nextscnu \\
& = h(\tilde\bfX_1, \tilde\bfX_2,\cdots,\tilde\bfX_{M+1}, \bfY_1,\bfY_2)
  - h(\tilde\bfX_1, \tilde\bfX_2,\cdots,\tilde\bfX_{M+1}| \bfY_1,\bfY_2)-h(\bfY_2)
  + n \nextscnu \\
& \le h(\tilde\bfX_1, \tilde\bfX_2,\cdots,\tilde\bfX_{M+1}, \bfY_1,\bfY_2)
    - h(\tilde\bfX_1, \tilde\bfX_2,\cdots,\tilde\bfX_{M+1}|
       \bfY_1,\bfY_2,\bfX_1,\bfX_2,\cdots,\bfX_{M+1})\nl
&\quad -h(\bfY_2) + n \nextscnu \\
& \le h(\tilde\bfX_1, \tilde\bfX_2,\cdots,\tilde\bfX_{M+1}, \bfY_1,\bfY_2)
    - h(\tilde\bfN_1, \tilde\bfN_2,\cdots,\tilde\bfN_{M+1}|
    \bfY_1,\bfY_2,\bfX_1,\bfX_2,\cdots,\bfX_{M+1})\nl
&\quad -h(\bfY_2) + n \nextscnu \\
& \le h(\tilde\bfX_1, \tilde\bfX_2,\cdots,\tilde\bfX_{M+1}, \bfY_1,\bfY_2)
    - h(\tilde\bfN_1, \tilde\bfN_2,\cdots,\tilde\bfN_{M+1})  -h(\bfY_2) + n \nextscnu \\
& \le h(\tilde\bfX_1, \tilde\bfX_2,\cdots,\tilde\bfX_{M+1}, \bfY_1,\bfY_2)
 -h(\bfY_2) + n \nextsc \\
& = h(\tilde\bfX_1, \tilde\bfX_2,\cdots,\tilde\bfX_{M+1})
+h(\bfY_1,\bfY_2|\tilde\bfX_1, \tilde\bfX_2,\cdots,\tilde\bfX_{M+1})
-h(\bfY_2) + n \nextscnu \\
& \le h(\tilde\bfX_1, \tilde\bfX_2,\cdots,\tilde\bfX_{M+1})  -h(\bfY_2) + n \nextsc
\label{eqn:gwch_general_ub_for_m_helpers_reconstruction}
\\
& = \sum_{i=1}^{M+1}h(\tilde\bfX_i)   -h(\bfY_2) + n \nextscnu
\end{align}
where \eqn{eqn:gwch_general_ub_for_m_helpers_reconstruction} is due to
$h(\bfY_1,\bfY_2|\tilde\bfX_1, \tilde\bfX_2,\cdots,\tilde\bfX_{M+1}) \le n
\nextsc$. The intuition behind this is that, given all (slightly noisy versions of) the channel inputs,
(at high SNR) the channel outputs can be \emph{reconstructed}. To show this formally, we have
\begin{align}
& h(\bfY_1,\bfY_2 |\tilde\bfX_1, \tilde\bfX_2,\cdots,\tilde\bfX_{M+1}) \nl
&\quad\quad \le h(\bfY_1 |\tilde\bfX_1, \tilde\bfX_2,\cdots,\tilde\bfX_{M+1})
         + h(\bfY_2 |\tilde\bfX_1, \tilde\bfX_2,\cdots,\tilde\bfX_{M+1}) 
         \label{eqn:gwch_reconstruction_start}
         \\
&\quad\quad =   h\left(\sum_{i=1}^{M+1} h_i (\tilde\bfX_i - \tilde\bfN_i) + \bfN_1 \Bigg|
                  \tilde\bfX_1, \tilde\bfX_2,\cdots,\tilde\bfX_{M+1}\right) \nl
&\quad\quad\quad
         + h\left(\sum_{i=1}^{M+1} g_i (\tilde\bfX_i - \tilde\bfN_i) + \bfN_2 \Bigg|
                  \tilde\bfX_1, \tilde\bfX_2,\cdots,\tilde\bfX_{M+1}\right) \\
&\quad\quad =   h\left(- \sum_{i=1}^{M+1} h_i  \tilde\bfN_i + \bfN_1 \Bigg|
                  \tilde\bfX_1, \tilde\bfX_2,\cdots,\tilde\bfX_{M+1}\right) \nl
&\quad\quad\quad
         + h\left(- \sum_{i=1}^{M+1} g_i  \tilde\bfN_i + \bfN_2 \Bigg|
                  \tilde\bfX_1, \tilde\bfX_2,\cdots,\tilde\bfX_{M+1}\right) \\
&\quad\quad \le h\left(- \sum_{i=1}^{M+1} h_i  \tilde\bfN_i + \bfN_1 \right)
         + h\left(- \sum_{i=1}^{M+1} g_i  \tilde\bfN_i + \bfN_2 \right) \\
&\quad\quad \stackrel{\triangle}{=} n \nextscnu
         \label{eqn:gwch_reconstruction_end}
\end{align}
which completes the proof of \eqn{eqn:gwch_general_ub_for_m_helpers}.

Finally, we show \eqn{eqn:gwch_general_ub_for_m_helpers_kill_y2}. To this end,
fixing a $j$, which could be arbitrary, we express $\bfY_2$ in a
stochastically equivalent form
$\tilde\bfY_2$, i.e.,
\begin{align}
\bfY_2       & = g_j \bfX_j + \sum_{i=1,i\neq j}^{M+1}g_i \bfX_i + \bfN_2 
    \label{eqn:gwch_kill_y2_start}    
\\
\tilde\bfY_2 & = g_j \tilde\bfX_j + \sum_{i=1,i\neq j}^{M+1}g_i \bfX_i + \bfN_2'
\end{align}
have  the same distribution, where $\bfN_2'$ is an i.i.d.~sequence of a random
variable $N_2'$ which is Gaussian with zero-mean and variance $(1-g_j^2\tilde\sigma_j^2)$, and
is independent of all other random variables. Then, we have
\begin{align}
h(\bfY_2)
 & = h(\tilde\bfY_2) \\
 & = h\left(g_j \tilde\bfX_j + \sum_{i=1,i\neq j}^{M+1}g_i \bfX_i + \bfN_2'\right) \\
 & \ge h\left(g_j \tilde\bfX_j\right) \label{eqn:added-ref-conv-lemma1}\\
 & = n \log \left|{g_j}\right| +  h(\tilde\bfX_j)
\label{eqn:gwch_kill_y2_end}    
\end{align}
where \eqn{eqn:added-ref-conv-lemma1} is due to the differential entropy version of \cite[Problem
2.14]{cover_it_book}. Substituting this into
\eqn{eqn:gwch_general_ub_for_m_helpers} gives us
\eqn{eqn:gwch_general_ub_for_m_helpers_kill_y2}.
\end{Proof}

\subsection{Role of a Helper}
\label{sec:gwch_general_role_of_helper}

Intuitively, a cooperative jamming signal from a helper may potentially increase
the secrecy of the legitimate transmitter-receiver pair by creating extra
equivocation at the eavesdropper. However, if the helper creates
too much equivocation, it may also hurt the decoding performance of the
legitimate receiver. Since the legitimate receiver needs to decode message
$W$ by observing $\bfY_1$, there must exist a constraint on the cooperative
jamming signal of the helper. To this end, we develop a constraint on the
differential entropy of (the noisy version of) the cooperative jamming signal of any given helper,
helper $j$ in \eqn{eqn:gwch_general_ub_for_helper}, in terms of the differential entropy of
the legitimate user's channel output and the message rate $H(W)$, in the following lemma.
The inequality in this lemma, \eqn{eqn:gwch_general_ub_for_helper}, can
alternatively be interpreted as an upper bound on the message rate, i.e., on
$H(W)$, in terms of the difference of the differential entropies of the channel
output of the legitimate receiver and the channel input of the $j$th helper; in
particular, the higher the differential entropy of the cooperative jamming
signal the lower this upper bound will be. This motivates not using i.i.d.~Gaussian
cooperative jamming signals which have the highest differential entropy.

Finally, we note as an aside that,
since this upper bound is derived based on the reliability of the legitimate
user's decoding (not involving any secrecy constraints), it can be used in \dof
calculations in settings not involving secrecy. We show an application of this
lemma in a non-secrecy context by developing an alternative proof for the
multiplexing gain of the $K$-user Gaussian interference channel, which was
originally proved in \cite{multiplexing_gain_of_networks}, in Appendix
\ref{sec:alternative_proof_of_mg_k_gic}.

\begin{lemma}
\label{lemma:gwch_general_ub_for_helper}
For reliable decoding at the legitimate receiver, the differential entropy of the input signal of helper $j$, $\bfX_j$, must satisfy
\begin{equation}
 h(\bfX_j + \tilde\bfN)\le  h(\bfY_1)  - H(W) + n {c}
  \label{eqn:gwch_general_ub_for_helper}
\end{equation}
where $c$ is a constant which does not depend on $P$, and $\tN$ is a new
Gaussian noise independent of all other random variables with $\sigma_{\tN}^2 < \frac{1}{h_j^2}$, and $\tilde\bfN$ is an i.i.d.~sequence of $\tilde N$.
\end{lemma}

\begin{Proof}
To reliably decode the message at the legitimate receiver, we must have
\begin{align}
nR=H(W) & \le I( \bfX_1;\bfY_1) \\
& = h(\bfY_1) - h(\bfY_1 |  \bfX_1) \\
& = h(\bfY_1) - h\left(\sum_{i=2}^{M+1} h_i \bfX_i + \bfN_1\right) \\
& \le h(\bfY_1) - h\left( h_j \bfX_j + \bfN_1\right)
\label{eqn:gwch_general_one_helper_less_than_m_helpers}
\\
& \le h(\bfY_1) - h\left( h_j \bfX_j + h_j \tilde\bfN\right)
\label{eqn:gwch_general_one_helper_n1_to_tildeN}
\\
& = h(\bfY_1) - h\left(  \bfX_j + \tilde\bfN\right) + n c
\end{align}
where  \eqn{eqn:gwch_general_one_helper_less_than_m_helpers} and
\eqn{eqn:gwch_general_one_helper_n1_to_tildeN} are due to the differential
entropy version of \cite[Problem 2.14]{cover_it_book}. In going from
\eqn{eqn:gwch_general_one_helper_less_than_m_helpers} to
\eqn{eqn:gwch_general_one_helper_n1_to_tildeN}, we also used the infinite
divisibility of Gaussian distribution and expressed $\bfN_1$ in its
stochastically equivalent form as $\bfN_1=h_j\tilde\bfN+\bfN'$ where $\bfN'$ is an i.i.d.~sequence of random variable $N'$ which is Gaussian with zero-mean and appropriate variance, and which is independent of all other random variables.
\end{Proof}
Note that, although we develop the inequality in
\eqn{eqn:gwch_general_ub_for_helper} for the message of transmitter-receiver
pair $1$, this result also holds for the message of any transmitter-receiver
pair in a multiple-message setting provided that the zero-mean Gaussian noise
$\tilde N$ has an appropriately small variance.

\section{Wiretap Channel with One Helper}
\label{sec:gwch_one_helper}

In this section, we consider the Gaussian wiretap channel with one helper as
formulated in Section~\ref{sec:gwch_model_gwch} for the case $M=1$. In this
section, we will show that the secure \dof is $\frac{1}{2}$ for almost all
channel gains as stated in the following theorem. The converse follows from
the general secrecy penalty upper bound in Section~\ref{sec:gwch_general_ub} and
the cooperative jamming signal upper bound in Section~\ref{sec:gwch_general_role_of_helper}.
The achievability is based on cooperative jamming with discrete signaling and real interference alignment.

\begin{theorem}
The secure \dof of the Gaussian wiretap channel with one helper is $\frac{1}{2}$ with probability one.
\end{theorem}

\subsection{Converse}
\label{sec:gwch_one_converse}

We start with \eqn{eqn:gwch_general_ub_for_m_helpers_kill_y2} of
Lemma~\ref{lemma:gwch_general_ub_for_m_helpers} with $M=1$ and by choosing $j=1$,
\begin{align}
n R
& \le  \sum_{i=1, i\neq j}^{M+1} h( \tilde\bfX_i) +  n c' \\
& =    h( \tilde\bfX_2) +  n c' \\
& \le  h(\bfY_1) - H(W) +  n \nextsc \label{eqn:proof_constant_2} \\
& \le \frac{n}{2} \log P-H(W)+ n \nextsc \label{eqn:proof_constant_2_next}
\end{align}
where \eqn{eqn:proof_constant_2} is due to
Lemma~\ref{lemma:gwch_general_ub_for_helper}. By noting $H(W)=nR$ and using
\eqn{eqn:sec-dof-defn}, \eqn{eqn:proof_constant_2_next} implies that
\begin{equation}
D_s \le \frac{1}{2}
\end{equation}
which concludes the converse part of the theorem.

\subsection{Achievable Scheme}

\begin{figure}[t]
\centering
\includegraphics[scale=0.8]{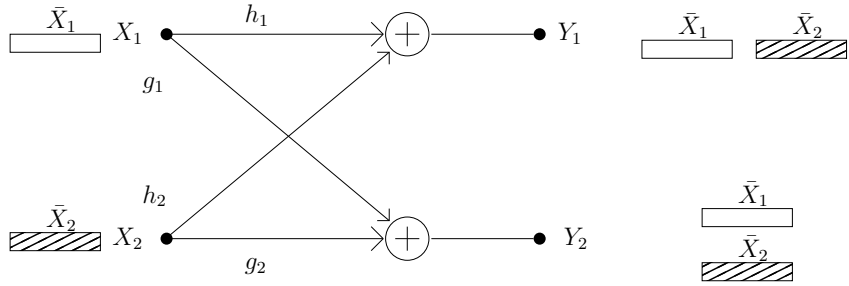}
\caption{Illustration of interference alignment for the Gaussian wiretap channel with one helper.}
\label{fig:gwc_one_helper_ia}
\end{figure}

To show the achievability by interference alignment, we slightly change the notation. Let $\etX_1\defn g_1 X_1 $, $\etX_2 \defn g_2 X_2$, $\alpha\defn h_1/g_1$, and $\beta\defn h_2/g_2$. Then, the channel model becomes
\begin{align}
\label{eqn:wiretap_one_helper_channel_model_equivalent_1}
Y_1 & =\alpha \etX_1 + \beta \etX_2 + N_1 \\
Y_2 & = \etX_1 + \etX_2 + N_2
\label{eqn:wiretap_one_helper_channel_model_equivalent_2}
\end{align}
Here $\etX_1$ is the input signal carrying the message $W$ of the legitimate transmitter and $\etX_2$ is the cooperative jamming signal from the helper. Our goal is to properly design $\etX_1$ and $\etX_2$ such that they are distinguishable at the legitimate receiver, meanwhile they align together at the eavesdropper. To prevent decoding of the message signal at the eavesdropper, we need to make sure that the cooperative jamming signal occupies the same \emph{dimensions} as the message signal at the eavesdropper; on the other hand, we need to make sure that the legitimate receiver is able to decode $\etX_2$, which in fact, is not useful. Intuitively, secrecy penalty is almost \emph{half} of the signal space, and we should be able to have a secure \dof of $\frac{1}{2}$. This is illustrated in Figure~\ref{fig:gwc_one_helper_ia}, and proved formally in the sequel.

We choose both of the input symbols $\etX_1$ and $\etX_2$ independent and uniformly distributed over the same PAM constellation
\begin{equation}
C(a,Q) = a \{ -Q, -Q+1, \ldots, Q-1,Q\}
\end{equation}
where $Q$ is a positive integer and $a$ is a real number used to normalize the transmission power, and is also the minimum distance between the points belonging to $C(a,Q)$.

Since $\bar{\bfX}_2$ is an i.i.d.~sequence and is independent of $\bar{\bfX}_1$, the following secrecy rate is always achievable \cite{wyner}
\begin{equation}
C_s \ge I(\etX_1;Y_1) - I(\etX_1;Y_2)
\end{equation}
In order to show that $D_s\ge \frac{1}{2}$, it suffices to prove that this lower bound provides $\frac{1}{2}$ secure d.o.f. To this end, we need to find a lower bound for $I(\etX_1;Y_1)$ and an upper bound for $I(\etX_1;Y_2)$. It is clear that
\begin{equation}
H(\etX_1) = H(\etX_2) = \log |C(a,Q)| = \log (2Q+1)
\end{equation}
Also, note that, besides the additive Gaussian noise, the observation at receiver $1$ is a linear combination of $\etX_1$ and $\etX_2$, i.e.,
\begin{equation}
Y_1 - N_1 = \alpha \etX_1 + \beta \etX_2
\end{equation}
where $\alpha$ and $\beta$ are rationally independent real numbers\footnote{ $a_1, a_2,\ldots, a_L$ are rationally independent if whenever $q_1,q_2,\ldots,q_L$ are rational numbers  then $\sum^L_{i=1} q_i a_i =0$ implies $q_i=0$ for all $i$.} with probability 1.

The space observed at receiver $1$ consists of $(2 Q+1)^2$ signal points. By using the Khintchine-Groshev theorem of Diophantine approximation in number theory, references \cite{real_inter_align_exploit,real_inter_align} bounded the minimum distance $d_{min}$ between the points in receiver 1's constellation as follows: For
any $\delta>0$, there exists a constant $k_\delta$ such that
\begin{equation}
\label{eqn:lb_of_d}
d_{min} \ge \frac{ k_\delta  a}{Q^{1+\delta}}
\end{equation}
for almost all rationally independent $\{\alpha,\beta\}$, except for a set of Lebesgue measure zero. Then, we can upper bound the probability of decoding error of such a PAM scheme by considering the additive Gaussian noise at receiver $1$ as follows,
\begin{align}
\pe\left[\etX_1\neq\hat{X}_1 \right] & \le \exp\left(  - \frac{d_{min}^2}{8}\right)
 \le \exp\left(-\frac{a^2k_\delta^2}{8 Q^{2(1+\delta)}}\right)
\end{align}
where $\hat{X}_1$ is the estimate for $\bar X_1$ obtained by choosing the closest point in the constellation based on observation $Y_1$. For any $\delta>0$, if we choose $Q = P^{\frac{1-\delta}{2(2+\delta)}}$ and $a=\gamma P^{\frac{1}{2}}/Q$, where $\gamma$ is a constant independent of $P$, then
\begin{align}
\pe\left[\etX_1\neq\hat{X}_1\right]
& \le \exp\left( -\frac{k_\delta^2 \gamma^2 P}{8 Q^{2(1+\delta)+2}} \right)
 = \exp\left( -\frac{k_\delta^2 \gamma^2 P^\delta}{8} \right)
\end{align}
and we can have $\pe\left[\etX_1\neq\hat{X}_1\right]\to 0$  as $P\to\infty$. To satisfy the power constraint at the transmitters, we can simply choose $\gamma \le \min(|g_1|,|g_2|)$. By Fano's inequality and the Markov chain $\etX_1\rightarrow Y_1\rightarrow\hat{X}_1$, we know that
\begin{align}
H(\etX_1 | Y_1)
& \le H(\etX_1|\hat{X}_1) \\
& \le 1 +\exp\left( -\frac{k_\delta^2 \gamma^2 P^\delta}{8}\right) \log(2Q+1)
\end{align}
which means that
\begin{align}
I(\etX_1;Y_1) & = H(\etX_1) - H(\etX_1|Y_1) \\
& \ge\left[ 1-\exp\left( -\frac{k_\delta^2 \gamma^2 P^\delta}{8}\right) \right] \log(2Q+1)   -1
\label{eqn:wiretap_one_helper_lb_ixy1}
\end{align}
On the other hand,
\begin{align}
I(\etX_1;Y_2)
& \le I(\etX_1;\etX_1 + \etX_2) \\
&  =  H(\etX_1 + \etX_2) - H(\etX_2 | \etX_1) \\
&  =  H(\etX_1 + \etX_2) - H(\etX_2 ) \\
& \le \log( 4Q +1) - \log( 2Q +1 )
\label{eqn:ub_for_combined_constellation} \\
& \le \log\frac{ 4Q +1 } { 2Q +1 } \\
& \le 1
\label{eqn:wiretap_one_helper_lb_ixy2}
\end{align}
where \eqn{eqn:ub_for_combined_constellation} is due to the fact that entropy of the sum $\etX_1 +\etX_2$ is maximized by the uniform distribution which takes values over a set of cardinality $4Q +1$.

Combining \eqn{eqn:wiretap_one_helper_lb_ixy1} and \eqn{eqn:wiretap_one_helper_lb_ixy2}, we have
\begin{align}
C_s
& \ge I(\etX_1;Y_1) - I(\etX_1;Y_2) \\
& \ge
\left[ 1-
\exp\left( -\frac{k_\delta^2 \gamma^2 P^\delta}{8}
\right) \right] \log(2Q+1)   -2 \\
& =
\left[ 1-
\exp\left( -\frac{k_\delta^2 \gamma^2 P^\delta}{8}
\right) \right] \log\left(2
P^{\frac{1-\delta}{2(2+\delta)}}
+1 \right)   -2 \\
& =\frac{1-\delta}{(2+\delta)} \left( \frac{1}{2} \log P \right) + o(\log P)
\end{align}
where the $o(\cdot)$ is the little-$o$ function. If we choose $\delta$ arbitrarily small, then we can achieve $\frac{1}{2}$ secure d.o.f., which concludes the achievability part of the theorem.

\section{Wiretap Channel with $M$ Helpers}
\label{sec:gwch_m_helper}

In this section, we consider the Gaussian wiretap channel with $M$ helpers as
formulated in Section~\ref{sec:gwch_model_gwch} for general $M>1$. In this
section, we will show that the secure \dof is $\frac{M}{M+1}$ for almost all
channel gains as stated in the following theorem. This shows that even though
the helpers are independent, the secure \dof increases monotonically with the
number of helpers $M$. The converse follows from the general secrecy
penalty upper bound in Section~\ref{sec:gwch_general_ub} and the cooperative
jamming signal upper bound in Section~\ref{sec:gwch_general_role_of_helper}. The
achievability is based on cooperative jamming of $M$ helpers with discrete
signaling and real interference alignment.

\begin{theorem}
The secure \dof of the Gaussian wiretap channel with $M$ helpers is $\frac{M}{M+1}$ with probability one.
\end{theorem}

\subsection{Converse}
We again start with \eqn{eqn:gwch_general_ub_for_m_helpers_kill_y2} of
Lemma~\ref{lemma:gwch_general_ub_for_m_helpers} with the selection of $j=1$
\begin{align}
n R
& \le\sum_{i=1,i\neq j}^{M+1} h( \tilde\bfX_i) +  n c' \\
& = \sum_{i=2}^{M+1} h( \tilde\bfX_i) +  n c' \\
& \le  M [ h(\bfY_1) -H(W)]  +  n \nextsc
\label{eqn:proof_m_helpers_next}
\end{align}
where \eqn{eqn:proof_m_helpers_next} is due to
Lemma~\ref{lemma:gwch_general_ub_for_helper} for each jamming signal
$\tilde\bfX_i$, $i=2,3,\cdots,M+1$. By noting $H(W)=nR$, \eqn{eqn:proof_m_helpers_next} implies that
\begin{align}
(M+1) n R & \le  M h(\bfY_1)  + n \nextscnu \label{eqn:gwch_ub_for_m_helpers} \\
& \le M \left( \frac{n}{2} \log P \right) + n \nextsc
\end{align}
which further implies from \eqn{eqn:sec-dof-defn}  that
\begin{equation}
D_s \le \frac{M}{M+1}
\end{equation}
which concludes the converse part of the theorem.

\subsection{Achievable Scheme}

\begin{figure}[t]
\centering
\includegraphics[scale=0.7]{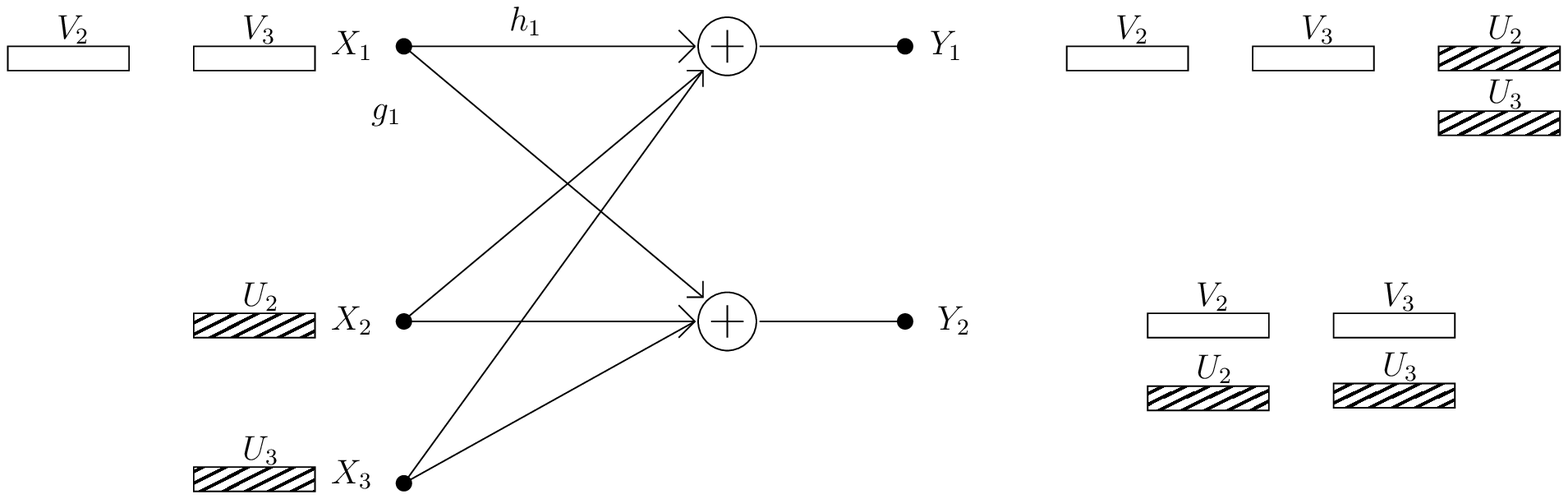}
\caption{Illustration of interference alignment for the Gaussian wiretap channel with M helpers. Here, $M=2$.}
\label{fig:gwc_m_helper_ia}
\end{figure}

Let $\{V_2,V_3,\cdots,V_{M+1},U_2,U_3,\cdots,U_{M+1}\}$ be mutually independent
discrete random variables, each of which  uniformly drawn from the same PAM
constellation $C(a,Q)$, where $a$ and $Q$ will be specified later. We choose the input signal of the legitimate transmitter as
\begin{equation}
X_1  = \sum_{k=2}^{M+1} \frac{g_k}{g_1 h_k}  V_k
\end{equation}
and the input signal of the $j$th helper, $j=2,3,\cdots,M+1$, as
\begin{equation}
X_j = \frac{1}{h_j} U_j
\end{equation}
Then, the observations of the receivers are
\begin{align}
Y_1
& = \sum_{k=2}^{M+1} \frac{h_1 g_k}{g_1 h_k} V_k+ \left( \sum_{j=2}^{M+1} U_j \right)+ N_1 \\
Y_2& = \sum_{k=2}^{M+1} \frac{g_k}{ h_k} \Big( V_k + U_k \Big)+ N_2
\end{align}
The intuition here is as follows. We use $M$ independent sub-signals $V_k$, $k=2,3,\cdots,M+1$, to represent the original message $W$. The input signal $X_1$ is a linear combination of $V_k$s. To cooperatively jam the eavesdropper, each helper $k$ aligns the cooperative jamming signal $U_k$ in the same \emph{dimension} as the sub-signal $V_k$ at the eavesdropper. At the legitimate receiver, all of the cooperative jamming signals $U_k$s are well-aligned such that they occupy a  small portion of the signal space. Since, with probability one, $\left\{1,\frac{h_1 g_2}{g_1 h_2}, \frac{h_1 g_3}{g_1 h_3}, \cdots, \frac{h_1 g_{M+1}}{g_1 h_{M+1}}\right\}$ are rationally independent, the signals $\left\{V_2,V_3,\cdots,V_{M+1}, \sum_{j=2}^{M+1} U_j \right\}$ can be distinguished by the legitimate receiver. As an example, the case of $M=2$ is shown in Figure~\ref{fig:gwc_m_helper_ia}.

Since, for each $j\neq 1$, ${\bfX}_j$ is an i.i.d.~sequence and independent of ${\bfX}_1$, the following
secrecy rate is achievable \cite{wyner}
\begin{equation}
C_s \ge I(X_1;Y_1) - I(X_1;Y_2)
\end{equation}

Now, we first bound the probability of decoding error. Note that the \emph{space} observed at receiver $1$ consists of $(2Q+1)^M (2MQ+1)$ points in $M+1$ \emph{dimensions}, and the sub-signal in each \emph{dimension} is drawn from a constellation of $C(a,MQ)$. Here, we use the property that $C(a,Q)\subset C(a,MQ)$.  By using the Khintchine-Groshev theorem of Diophantine approximation in number theory, we can bound the minimum distance $d_{min}$ between the points in receiver 1's \emph{space} as follows: For any $\delta>0$, there exists a constant $k_\delta$ such that
\begin{equation}
\label{eqn:gwch_lb_of_d_m_helper}
d_{min} \ge \frac{ k_\delta  a}{(MQ)^{M+\delta}}
\end{equation}
for almost all rationally independent $\left\{1,\frac{h_1 g_2}{g_1 h_2}, \frac{h_1 g_3}{g_1 h_3}, \cdots, \frac{h_1 g_{M+1}}{g_1 h_{M+1}}\right\}$, except for a set of Lebesgue measure zero. Then, we can upper bound the probability of decoding error of such a PAM scheme by considering the additive Gaussian noise at receiver $1$,
\begin{align}
\pe\left[X_1\neq\hat{X}_1 \right] & \le \exp\left(  - \frac{d_{min}^2}{8}\right)  
 \le \exp\left(  -
\frac{a^2k_\delta^2}{8 (MQ)^{2(M+\delta)}}\right)
\end{align}
where $\hat{X}_1$ is the estimate of $X_1$ by choosing the closest point in the constellation based on observation $Y_1$. For any $\delta>0$, if we choose $Q = P^{\frac{1-\delta}{2(M+1+\delta)}}$ and $a=\gamma P^{\frac{1}{2}}/Q$, where $\gamma$ is a constant independent of $P$, then
\begin{align}
\pe\left[X_1\neq\hat{X}_1\right]
& \le \exp\left( -\frac{k_\delta^2 \gamma^2 M^2 P}{8 (MQ)^{2(M+\delta)+2}} \right)  
 = \exp\left( -\frac{k_\delta^2 \gamma^2 M^2 P^\delta}{8 M^{2(M+1+\delta)}} \right)
\end{align}
and we can have $\pe\left[X_1\neq\hat{X}_1\right] \to 0$  as $P\to\infty$. To satisfy the power constraint at the transmitters, we can simply choose $\gamma \le \min([\sum_{k=2}^{M+1} (\frac{g_k}{g_1 h_k})^2]^{-1/2},|h_2|, |h_3|, \cdots, |h_{M+1}|)$. By Fano's inequality and the Markov chain $X_1\rightarrow Y_1\rightarrow\hat{X}_1$, we know that
\begin{align}
H(X_1 | Y_1)
& \le H(X_1|\hat{X}_1) \\
& \le 1 +
\exp\left( -\frac{k_\delta^2 \gamma^2 M^2 P^\delta}{8 M^{2(M+1+\delta)}}
\right) \log(2Q+1)^M
\end{align}
which means that
\begin{align}
I(X_1;Y_1) & = H(X_1) - H(X_1|Y_1) \\
& \ge
\left[ 1-
\exp\left(  -\frac{k_\delta^2 \gamma^2 M^2 P^\delta}{8 M^{2(M+1+\delta)}}
\right) \right] \log(2Q+1)^M  -1
\label{eqn:gwch_wiretap_m_helper_lb_ixy1}
\end{align}
On the other hand,
\begin{align}
  I(X_1;Y_2)
& \le I\left(X_1;\sum_{k=2}^{M+1} \frac{g_k}{ h_k} ( V_k + U_k )\right) \\
&  =  H\left( \sum_{k=2}^{M+1} \frac{g_k}{ h_k} ( V_k + U_k )\right)
- H\left( \sum_{k=2}^{M+1} \frac{g_k}{ h_k} ( V_k + U_k )\Big|X_1\right) \\
&  =  H\left( \sum_{k=2}^{M+1} \frac{g_k}{ h_k} ( V_k + U_k )\right)- H\left( \sum_{k=2}^{M+1} \frac{g_k}{ h_k} U_k\right) \\
& \le \log( 4Q +1)^M - \log( 2Q +1 )^M
\label{eqn:gwch_ub_for_combined_constellation_m_streams} \\
& \le M \log\frac{ 4Q +1 } { 2Q +1 } \\
& \le M
\label{eqn:gwch_wiretap_m_helper_lb_ixy2}
\end{align}
where \eqn{eqn:gwch_ub_for_combined_constellation_m_streams} is due to the fact that entropy of the sum $\sum_{k=2}^{M+1} \frac{g_k}{ h_k} ( V_k + U_k )$ is maximized by the uniform distribution which takes values over a set of cardinality $(4Q +1)^M$.

Combining \eqn{eqn:gwch_wiretap_m_helper_lb_ixy1} and \eqn{eqn:gwch_wiretap_m_helper_lb_ixy2}, we have
\begin{align}
C_s
& \ge I(X_1;Y_1) - I(X_1;Y_2) \\
& \ge \left[ 1-\exp\left(  -\frac{k_\delta^2 \gamma^2 M^2 P^\delta}{8 M^{2(M+1+\delta)}}
\right) \right] \log(2Q+1)^M  -(M+1) \\
& \ge \left[ 1-\exp\left(  -\frac{k_\delta^2 \gamma^2 M^2 P^\delta}{8 M^{2(M+1+\delta)}}
\right) \right] \log(2P^{\frac{1-\delta}{2(M+1+\delta)}}+1)^M  -(M+1) \\
& ={\frac{M(1-\delta)}{(M+1+\delta)}} \left(\frac{1}{2}\log P\right) + o(\log P)
\end{align}
where $o(\cdot)$ is the little-$o$ function. If we choose $\delta$
arbitrarily small, then we can achieve $\frac{M}{M+1}$ secure d.o.f., which concludes the achievability part of the theorem.

\section{Broadcast Channel with Confidential Messages and $M$ Helpers}
\label{sec:gbc_cm_helpers}

In this section, we consider the Gaussian broadcast channel with confidential
messages and $M$ helpers formulated in Section~\ref{sec:gwch_model_bch}. When
there are no helpers, i.e., $M=0$, due to the degradedness of the underlying
Gaussian broadcast channel, one of the users (stronger) has the secrecy capacity which is
equal to the secrecy capacity of the Gaussian wiretap channel, and the other user
(weaker) has zero secrecy capacity. Therefore, for both users, the secure \dof
is zero, implying that the sum secure \dof of the system is zero. Therefore, we
consider the case $M\ge 1$. In this section, we will show that the sum secure \dof
is $1$ for any $M \ge 1$, as stated in the following theorem.

\begin{theorem}
The sum secure \dof of the Gaussian broadcast channel with confidential messages and $M\ge1$ helpers is $1$ with probability one.
\end{theorem}

\subsection{Converse}
An immediate upper bound for the secure \dof of this problem is $1$, i.e.,
$D_{s,\Sigma} \le 1$ for any $M$. This comes from the fact that the \dof for the
Gaussian broadcast channel without any secrecy constraints is $1$, and this constitutes an
upper for the sum secure \dof also.

\subsection{Achievable Scheme}
In the following, we will show that a sum secure \dof of $1$ can be achieved for the case of $M=1$.
Since the achievable scheme with a single helper achieves the upper bound $D_{s,\Sigma} \le 1$,
the sum secure \dof for all $M\ge 1$ is $1$. Therefore, if we have more than one helper, then all but one
helper may remain silent.

\begin{figure}[t]
\centering
\includegraphics[scale=0.8]{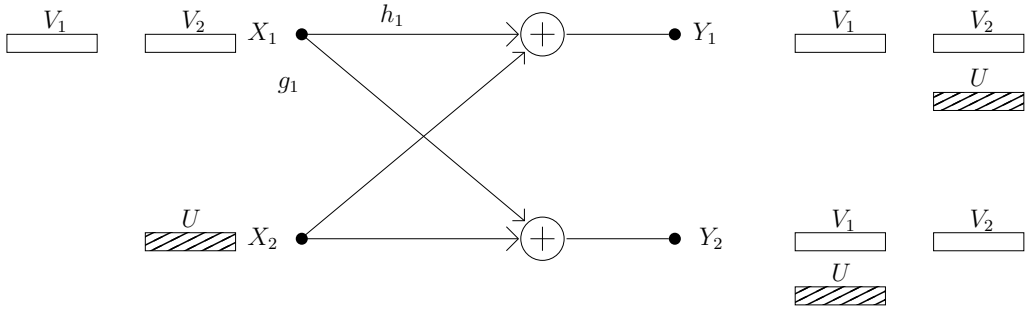}
\caption{Illustration of interference alignment for the Gaussian broadcast channel with confidential messages and one helper.}
\label{fig:gbc_one_helper_ia}
\end{figure}

We use the equivalent channel expression in
\eqn{eqn:wiretap_one_helper_channel_model_equivalent_1} and
\eqn{eqn:wiretap_one_helper_channel_model_equivalent_2}. Let $V_1,V_2$ and $U$
be three mutually independent random variables which are identically and
uniformly distributed over the constellation $C(a,Q)$, where $a$ and $Q$ will be
specified later. We assign channel inputs as $\etX_1 = V_1 +
\frac{\beta}{\alpha} V_2$ and $\etX_2 = U$. Then, the observations at the two
receivers are:
\begin{align}
Y_1 & = \alpha V_1 + \beta ( V_2 + U) + N_1 \\
Y_2 & = ( V_1 + U)  + \frac{\beta}{\alpha}  V_2  + N_2
\end{align}
We use two independent variables $V_1$ and $V_2$ to carry the messages $W_1$ and
$W_2$ that go to the two receivers. In order to ensure that the messages are
kept secure against the unintended receiver, we align the cooperative noise
signal $U$ from the helper in the \emph{dimension} of  $V_2$ at receiver $1$, and
in the \emph{dimension} of $V_1$ at receiver $2$. This is illustrated in
Figure~\ref{fig:gbc_one_helper_ia}.

Since $\bar{\bfX}_2$ is an i.i.d.~sequence, the following secrecy rate pair is achievable \cite[Theorem 4]{secrecy_ic}
\begin{align}
R_{1} & \ge I(V_1;Y_1) - I(V_1;Y_2|V_2) \\
R_{2} & \ge I(V_2;Y_2) - I(V_2;Y_1|V_1)
\end{align}

By using Khintchine-Groshev theorem, it is easy to verify that receiver $i$ can decode $V_i$, for $i=1,2$ with arbitrarily small probability of decoding error  with probability one, i.e., for any $\delta>0$, there exists a constant $k_\delta$ such that the minimum distance $d_{min}$ between points at receiver $i$ is,
\begin{equation}
d_{min} \ge \frac{ k_\delta  a}{(2Q)^{1+\delta}}
\end{equation}
for almost all rationally independent $\{\alpha,\beta\}$, except for a set of Lebesgue measure zero. Then, we can upper bound the probability of decoding error for such a PAM scheme by considering the additive Gaussian noise at receiver $i$ as,
\begin{align}
\pe\left[V_i\neq\hat{V}_i \right] & \le \exp\left(  - \frac{d_{min}^2}{8}\right)  
 \le \exp\left(  -\frac{a^2k_\delta^2}{8 (2Q)^{2(1+\delta)}}\right)
\end{align}
where $\hat{V}_i$ is the estimate for $V_i$ by choosing the closest point in the constellation based on observation $Y_i$. For any $\delta>0$, if $Q = P^{\frac{1-\delta}{2(2+\delta)}}$, $a=\gamma P^{\frac{1}{2}}/Q$, and $\gamma$ is a positive constant satisfying
\begin{equation}
\gamma \le \min\left\{ |g_1|{\left[1 +\left( \frac{\beta}{\alpha}\right)^2\right]}^{-1/2}, |g_2|\right\}
\end{equation}
then
\begin{align}
\pe\left[V_i\neq\hat{V}_i \right]
& \le \exp\left( -\frac{4 k_\delta^2 \gamma^2 P}{8 (2Q)^{2(2+\delta)}} \right)  
 = \exp\left( -\frac{ k_\delta^2 \gamma^2 P^\delta}{2^{2\delta+5}} \right)
\end{align}
and we can have $\pe\left[V_i\neq\hat{V}_i \right]\to 0$  as $P\to\infty$. By Fano's inequality and the Markov chain $V_i\rightarrow Y_i\rightarrow\hat{V}_i$, we
know that
\begin{align}
H(V_i | Y_i)
& \le H(V_i|\hat{V}_i) \\
& \le 1 + \exp\left(  -\frac{ k_\delta^2 \gamma^2 P^\delta}{2^{2\delta+5}}\right) \log(2Q+1)
\end{align}
which means that
\begin{align}
I(V_i;Y_i) & = H(V_i) - H(V_i|Y_i) \\
& \ge\left[ 1-\exp\left(  -\frac{ k_\delta^2 \gamma^2 P^\delta}{2^{2\delta+5}}\right) \right] \log(2Q+1)   -1 \\
& = \frac{1-\delta}{2+\delta} \left( \frac{1}{2} \log P\right)  + o(\log P)
\end{align}
for $i=1$ or $2$.

On the other hand, for $i=1$, we have
\begin{align}
I(V_1;Y_2|V_2)
& \le I\left(V_1; V_1 + U + \frac{\beta}{\alpha}  V_2 \Big| V_2\right) \\
& = H(V_1 +U) - H(U) \\
& \le 1
\end{align}
Similarly, for $i=2$, we have
\begin{align}
I(V_2;Y_1|V_1)
& \le I\left(V_2; \alpha V_1 + \beta ( V_2 + U) \Big| V_1\right) \\
& = H(V_2 +U) - H(U) \\
& \le 1
\end{align}
which implies that the following sum secrecy rate is achievable
\begin{equation}
R_{1} + R_{2} \ge \frac{2-2\delta}{2+\delta} \left( \frac{1}{2} \log P \right) + o(\log P)
\end{equation}
If we choose $\delta$ small enough, then we can have $D_{s,\Sigma} \ge 1$. Combining this with the upper bound $D_{s,\Sigma} \le 1$, we conclude that
\begin{equation}
D_{s,\Sigma} = 1
\end{equation}
with probability one.

\section{Two-User Interference Channel with Confidential Messages and No Helpers}
\label{sec:gic_cm}
In this section, we consider the two-user Gaussian interference channel with confidential messages formulated in Section~\ref{sec:gwch_model_ich} for the case of no helpers, i.e., $M=0$. The case of $M \ge 1$ will be presented in Section~\ref{sec:gic_cm_m_helpers}. For the case of no helpers, we show that the sum secure \dof is $\frac{2}{3}$ as stated in the following theorem.

\begin{theorem}
The sum secure \dof of the two-user Gaussian interference channel with confidential messages is $\frac{2}{3}$ with probability one.
\end{theorem}

\subsection{Converse}
We first start with \eqn{eqn:gwch_general_ub_for_m_helpers} of Lemma~\ref{lemma:gwch_general_ub_for_m_helpers} to upper bound the individual rate $R_1$ of message $W_1$
\begin{align}
n R_1
& \le  h( \tilde\bfX_1) + h( \tilde\bfX_2) - h(\bfY_2) +  n c\\
& \le  h( \tilde\bfX_1) + h( \bfY_1) - H(W_1) - h(\bfY_2) +  n \nextsc
\label{eqn:gic_converse_part_1}\\
& \le h(\bfY_2) - H(W_2) + h( \bfY_1) - H(W_1) - h(\bfY_2) +  n \nextsc
\label{eqn:gic_converse_part_2}
\end{align}
where \eqn{eqn:gic_converse_part_1} is due to applying
Lemma~\ref{lemma:gwch_general_ub_for_helper} for $h(\tilde\bfX_2)$ and
\eqn{eqn:gic_converse_part_2} is due to applying
Lemma~\ref{lemma:gwch_general_ub_for_helper} once again for $h(\tilde\bfX_1)$.
By noting that $H(W_1)=n R_1$ and $H(W_2)=n R_2$, from \eqn{eqn:gic_converse_part_2}, we have
\begin{equation}
2 n R_{1}  + n R_2 \le  h(\bfY_1)  +  n \nextscnu \label{eqn:intermediate-bound1}
\end{equation}
We use the same method to get a symmetric upper bound on the individual rate $R_2$ of message $W_2$ as
\begin{equation}
n R_{1}  + 2 n R_2 \le  h(\bfY_2)  +  n \nextsc \label{eqn:intermediate-bound2}
\end{equation}
Then, combining \eqn{eqn:intermediate-bound1} and \eqn{eqn:intermediate-bound2}, we get
\begin{align}
3  (n R_{1}  +n R_2) & \le h(\bfY_1) + h(\bfY_2)  +  n \nextsc \\
& \le 2  \left( \frac{n}{2} \log P \right) + n \nextsc
\end{align}
which means
\begin{equation}
D_{s,\Sigma} \le \frac{2}{3}
\end{equation}
which concludes the converse part of the theorem.

\subsection{Achievable Scheme}

\begin{figure}[t]
\centering
\includegraphics[scale=0.7]{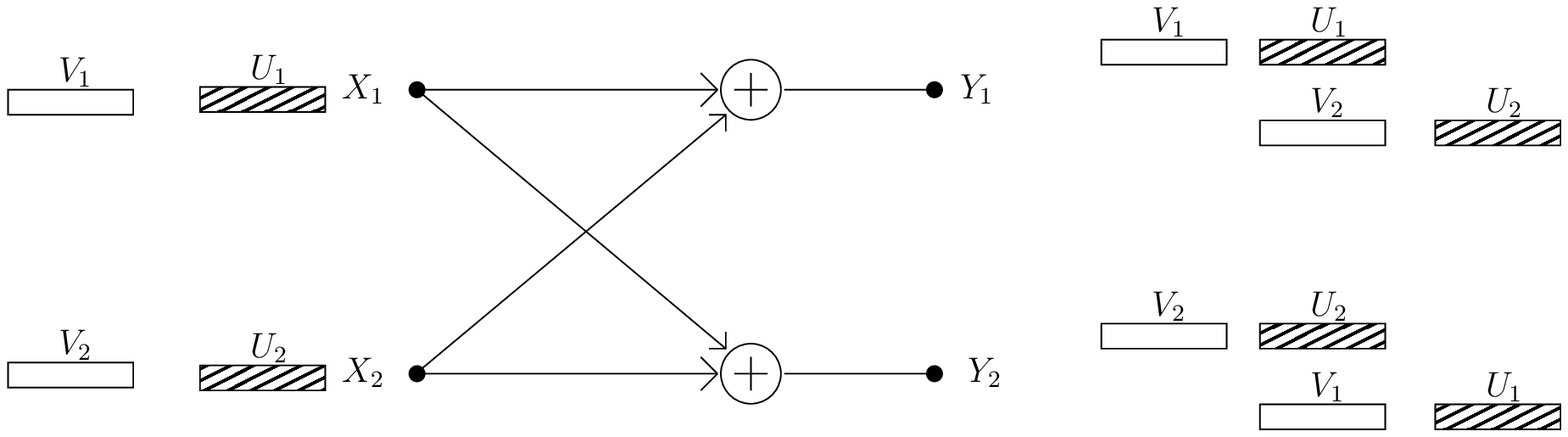}
\caption{Illustration of interference alignment for the two-user Gaussian interference channel with confidential messages (no helpers).}
\label{fig:gic_cm_ia}
\end{figure}

Let $\{V_1,U_1,V_2,U_2\}$ be mutually independent discrete random variables.
Each of them is uniformly and independently drawn from the same constellation
$C(a,Q)$, where $a$ and $Q$ will be specified later. Here, the role of $V_i$ is
to carry message $W_i$, and the role of $U_i$ is the cooperative jamming signal
to help the transmitter-receiver pair $j\neq i$. We choose the input signals of the
transmitters as:
\begin{align}
X_1 & = V_1 + \frac{h_{2,1}}{h_{1,1}} U_1 \\
X_2 & = V_2 + \frac{h_{1,2}}{h_{2,2}} U_2
\end{align}
With these input signal selections, observations of the receivers are
\begin{align}
Y_1& = h_{1,1}V_1 + {h_{2,1}} \big( U_1 + V_2 \big)
+ \frac{ h_{2,1} h_{1,2} }{ h_{2,2} } U_2 + N_1\\
Y_2& = h_{2,2}V_2 + {h_{1,2}} \big( U_2 +V_1 \big)
+ \frac{ h_{2,1} h_{1,2} }{ h_{1,1} } U_1 + N_2
\end{align}
Since, for each $i$ and $j\neq i$, $V_i$ and $U_i$ are not in the same
\emph{dimension} at both
receivers, we align $U_i$ in the \emph{dimension} of $V_j$ at receiver $i$ such
that $V_j$ is \emph{secure} and $V_i$ can occupy a \emph{larger} space. This is
illustrated in Figure~\ref{fig:gic_cm_ia}.

By \cite[Theorem 2]{secrecy_ic}, we know that the following secrecy rate pair is
achievable
\begin{align}
R_{1} & \ge I(V_1;Y_1) - I(V_1;Y_2|V_2) \\
R_{2} & \ge I(V_2;Y_2) - I(V_2;Y_1|V_1)
\end{align}
For receiver $1$, by using the Khintchine-Groshev theorem of Diophantine approximation in number theory, we can bound the minimum distance $d_{min}$ between points in the receiver's \emph{space}, i.e.,  for any $\delta>0$, there exists a constant $k_\delta$ such that
\begin{equation}
d_{min} \ge \frac{ k_\delta  a}{(2Q)^{2+\delta}}
\end{equation}
for almost all rationally independent $\left\{h_{1,1}, {h_{2,1}},  \frac{h_{2,1} h_{1,2}}{h_{2,2}}\right\}$, except for a set of Lebesgue measure zero. Then, we can upper bound the probability of decoding error of such a PAM scheme by considering the additive Gaussian noise at receiver $1$ as,
\begin{align}
\pe\left[V_1\neq\hat{V}_1 \right] & \le \exp\left(  - \frac{d_{min}^2}{8}\right)  
 \le \exp\left(  - \frac{a^2k_\delta^2}{8 (2Q)^{2(2+\delta)}}\right)
\end{align}
where $\hat{V}_1$ is the estimate of $V_1$ by choosing the closest point in the constellation based on observation $Y_1$. For any $\delta>0$, if we choose $Q = P^{\frac{1-\delta}{2(3+\delta)}}$ and $a=\gamma P^{\frac{1}{2}}/Q$, where
\begin{equation}
\gamma <\min_{i} \frac{1}{\sqrt{1+ \left(\frac{h_{j,i}}{h_{i,i}}\right)^2}}
\end{equation}
is a constant independent of $P$ to normalize the average power of the input signals. Then,
\begin{align}
\pe\left[V_1\neq\hat{V}_1\right]
& \le \exp\left( -\frac{k_\delta^2 \gamma^2 4 P}{8 (2Q)^{2(2+\delta)+2}} \right)  
= \exp\left( -\frac{k_\delta^2 \gamma^2  P^\delta}{ 2^{2\delta+7}} \right)
\end{align}
and we can have $\pe\left[V_1\neq\hat{V}_1\right]\to 0$  as $P\to\infty$.

To lower bound the achievable rate $R_1$, we first note that
\begin{align}
I(V_1;Y_1) & \ge I(V_1;\hat{V}_1) \\
& = H(V_1) - H(V_1|\hat{V}_1) \\
& \ge \left[1  - \exp\left( -\frac{k_\delta^2 \gamma^2  P^\delta}{
2^{2\delta+7}} \right)\right] \log(2Q+1) -1 \\
& = \frac{1-\delta}{3+\delta} \left( \frac{1}{2} \log P \right) + o (\log P) \label{eqn:step1}
\end{align}
On the other hand,
\begin{align}
I(V_1;Y_2|V_2) & \le I(V_1;Y_2,U_1|V_2) \\
& =  I(V_1;Y_2|V_2,U_1) \\
& \le I\left(V_1;   {h_{1,2}} ( U_2 +V_1 ) |V_2,U_1\right)\\
& = H( U_2 +V_1 ) - H(U_2) \\
& \le \log(4Q+1) - \log(2Q+1) \\
& \le 1 \label{eqn:step2}
\end{align}
Combining \eqn{eqn:step1} and \eqn{eqn:step2}, we obtain
\begin{align}
R_1 & \ge I(V_1;Y_1) - I(V_1;Y_2|V_2)  \\
& \ge \frac{1-\delta}{3+\delta} \left( \frac{1}{2} \log P \right) + o (\log P)
\end{align}
By applying this same analysis to rate $R_2$, we can obtain a symmetric result for $R_2$. Then, by choosing $\delta$ arbitrarily small, we can achieve $\frac{2}{3}$ sum secure d.o.f.

\section{Two-User Interference Channel with Confidential Messages and $M$ Helpers}
\label{sec:gic_cm_m_helpers}

In this section, we consider the two-user Gaussian interference channel with confidential messages formulated in Section~\ref{sec:gwch_model_ich} for the general case of $M\ge 1$ helpers. For this general case, we show that the sum secure \dof is $1$ as stated in the following theorem.

\begin{theorem}
The sum secure \dof of the two-user Gaussian interference channel with confidential messages and $M\ge 1$ helpers is $1$ with probability one.
\end{theorem}

\subsection{Converse}
An immediate upper bound for the secure \dof of this problem is $1$, i.e.,
$D_{s,\Sigma} \le 1$ for any $M$. This comes from the fact that the \dof for the two-user
interference channel without any secrecy constraints is $1$, and this constitutes an
upper for the sum secure \dof also. The fact that the \dof of the two-user interference channel
is $1$ was first proved in \cite{multiplexing_gain_of_networks}. We provide an alternative proof
to this fact using the techniques developed in this paper in Appendix~\ref{sec:alternative_proof_of_mg_k_gic}.

\subsection{Achievable Scheme}

In the following, we will show that a sum secure \dof of $1$ can be achieved for the case of $M=1$.
Since the achievable scheme with a single helper achieves the upper bound $D_{s,\Sigma} \le 1$,
the sum secure \dof for all $M\ge 1$ is $1$. Therefore, if we have more than one helpers, then all but one
helper may remain silent.

\begin{figure}[t]
\centering
\includegraphics[scale=0.8]{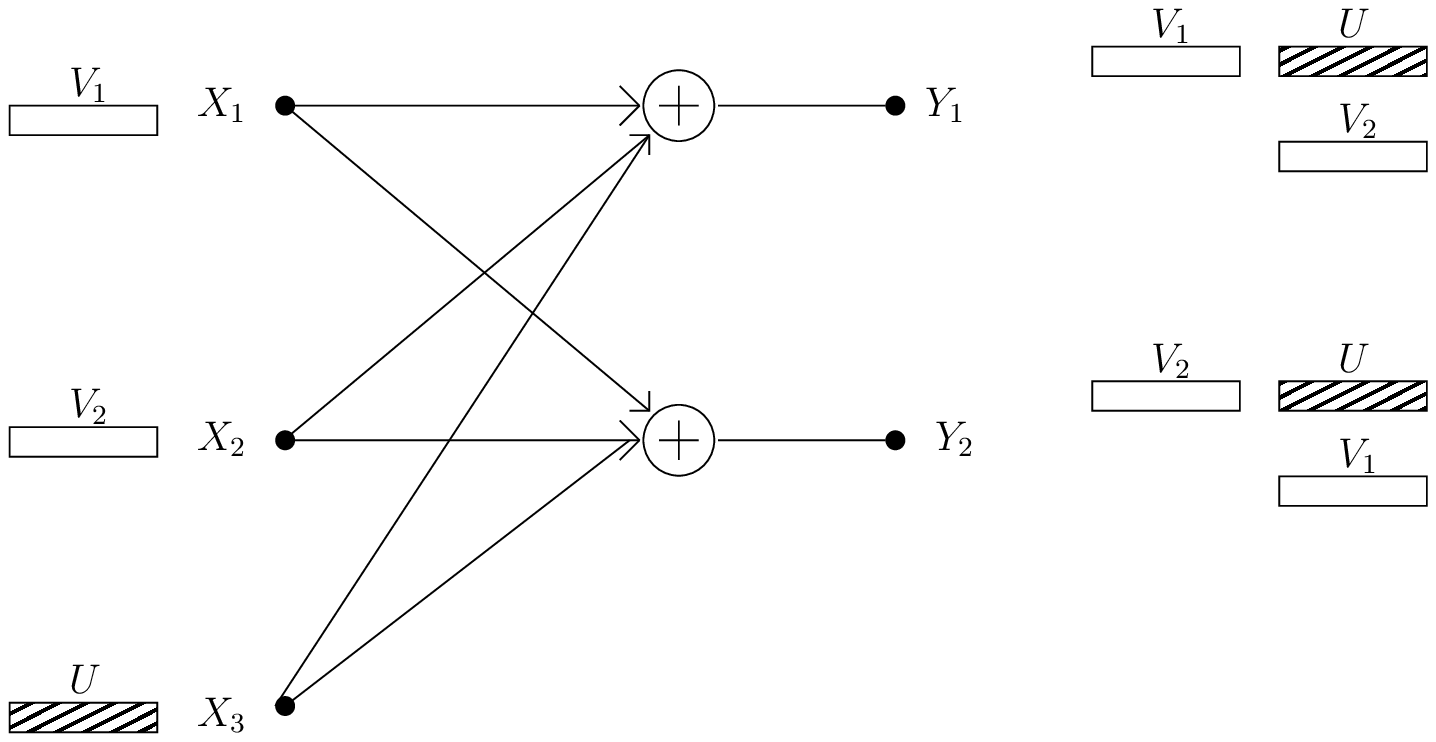}
\caption{Illustration of interference alignment for the two-user Gaussian
interference channel with confidential messages and one helper.}
\label{fig:gic_cm_m_helper_ia}
\end{figure}

Let $\{V_1,V_2,U\}$ be mutually independent discrete random variables. Each of
them is uniformly and independently drawn from the same constellation $C(a,Q)$,
where $a$ and $Q$ will be specified later. Here, the role of $V_i$ is to carry
message $W_i$, and the role of $U$ is the cooperative jamming signal from the
helper. We choose the input signals of the transmitters as:
\begin{align}
X_1 & = \frac{h_{3,2}}{h_{1,2}} V_1 \\
X_2 & = \frac{h_{3,1}}{h_{2,1}} V_2 \\
X_3 & = U
\end{align}
With these input signal selections, observations of the receivers are
\begin{align}
Y_1 & =  \frac{ h_{3,2} h_{1,1} }{ h_{1,2} }V_1 + {h_{3,1}} \big( U + V_2 \big)
+ N_1 \\
Y_2 & =  \frac{ h_{3,1} h_{2,2} }{ h_{2,1} }V_2 + {h_{3,2}} \big( U + V_1 \big)
+ N_2
\end{align}
For each $i$ and $j\neq i$, we align $U$ in the \emph{dimension} of $V_j$ at receiver $i$ such that $V_j$ is \emph{secure} and $V_i$ can be decoded. This is illustrated in Figure~\ref{fig:gic_cm_m_helper_ia}.

Since $\bfU$ is an i.i.d.~sequence, by \cite[Theorem 2]{secrecy_ic}, we know that the following secrecy rate pair is achievable
\begin{align}
R_{1} & \ge I(V_1;Y_1) - I(V_1;Y_2|V_2) \\
R_{2} & \ge I(V_2;Y_2) - I(V_2;Y_1|V_1)
\end{align}
For receiver $1$, by using the Khintchine-Groshev theorem of Diophantine approximation in number theory, we can bound the minimum distance $d_{min}$ between the points in receiver's \emph{space}, i.e.,  for any $\delta>0$, there exists a constant $k_\delta$ such that
\begin{equation}
d_{min} \ge \frac{ k_\delta  a}{(2Q)^{1+\delta}}
\end{equation}
for almost all rationally independent $\left\{\frac{ h_{3,2} h_{1,1} }{ h_{1,2} }, {h_{3,1}} \right\}$, except for a set of Lebesgue measure zero. Then, we can upper bound the probability of decoding error of such a PAM scheme by considering the additive Gaussian noise at receiver $1$ as,
\begin{align}
\pe\left[V_1\neq\hat{V}_1 \right]
& \le \exp\left(  - \frac{d_{min}^2}{8}\right)  
 \le \exp\left(  -
\frac{a^2k_\delta^2}{8 (2Q)^{2(1+\delta)}}\right)
\end{align}
where $\hat{V}_1$ is the estimate of $V_1$ by choosing the closest point in the constellation based on the observation $Y_1$. For any $\delta>0$, if we choose $Q = P^{\frac{1-\delta}{2(2+\delta)}}$ and $a=\gamma P^{\frac{1}{2}}/Q$, where
\begin{equation}
\gamma <\min \left(
\left|  \frac{ h_{3,2} }{ h_{1,2} }  \right|^ {-1},
\left|  \frac{ h_{3,1} }{ h_{2,1} }  \right|^ {-1},
1
\right)
\end{equation}
is a constant independent of $P$ to normalize the average power of the input signals. Then,
\begin{align}
\pe\left[V_1\neq\hat{V}_1\right]
& \le \exp\left( -\frac{k_\delta^2 \gamma^2 4 P}{8 (2Q)^{2(1+\delta)+2}} \right)  
 = \exp\left( -\frac{k_\delta^2 \gamma^2  P^\delta}{2^{2\delta+5}} \right)
\end{align}
and we can have $\pe\left[V_1\neq\hat{V}_1\right]\to 0$  as $P\to\infty$.

To lower bound the achievable rate $R_1$, we first note that
\begin{align}
I(V_1;Y_1) & \ge I(V_1;\hat{V}_1) \\
& = H(V_1) - H(V_1|\hat{V}_1) \\
& \ge \left[1  - \exp\left(
-\frac{k_\delta^2 \gamma^2  P^\delta}{2^{2\delta+5}} \right) \right]\log(2Q+1)
-1\\
& = \frac{1-\delta}{2+\delta} \left( \frac{1}{2} \log P \right) + o (\log P) \label{eqn:combine1}
\end{align}
On the other hand,
\begin{align}
I(V_1;Y_2|V_2)
& \le I\Big(V_1;   {h_{3,2}} ( U +V_1 )  |V_2\big)\\
& = H( U +V_1 ) - H(U) \\
& \le \log(4Q+1) - \log(2Q+1) \\
& \le 1 \label{eqn:combine2}
\end{align}
Combining \eqn{eqn:combine1} and \eqn{eqn:combine2}, we obtain
\begin{align}
R_1 & \ge I(V_1;Y_1) - I(V_1;Y_2|V_2)  \\
& \ge \frac{1-\delta}{2+\delta} \left( \frac{1}{2} \log P \right) + o (\log P)
\end{align}
By applying this same analysis to rate $R_2$, we can obtain a symmetric result for $R_2$. Then, by choosing $\delta$ arbitrarily small, we can achieve $1$ sum secure \dof with probability one for almost all channel gains for the $M=1$ case. 

\section{$K$-User Multiple Access Wiretap Channel}
\label{sec:gmacw_k}

In this section, we consider the $K$-user multiple access wiretap channel
formulated in Section~\ref{sec:gwch_model_macw}. We show that the sum secure
\dof of this channel is $\frac{K(K-1)}{K(K-1)+1}$ as stated in the following
theorem.

\begin{theorem}
The sum secure \dof of the $K$-user Gaussian multiple access wiretap channel is $\frac{K(K-1)}{K(K-1)+1}$ with probability one.
\end{theorem}

\subsection{Converse}
We start with the sum rate and derive an upper bound similar to Lemma~\ref{lemma:gwch_general_ub_for_m_helpers}
\begin{align}
n\sum_{i=1}^K R_i
& = \sum_{i=1}^K H(W_i) = H(W_1^K)\\
& \le I(W_1^K;\bfY_1,\bfY_2)  - I(W_1^K;\bfY_2) + n \nextscnu \\
& = I(W_1^K;\bfY_1|\bfY_2)  + n \nextscnu \\
& \le I(\bfX_1^K;\bfY_1|\bfY_2)  + n \nextscnu \\
& = h(\bfY_1|\bfY_2) - h(\bfY_1|\bfY_2,\bfX_1^K)  + n \nextscnu \\
& = h(\bfY_1|\bfY_2) - h(\bfN_1|\bfY_2,\bfX_1^K)  + n \nextscnu \\
& \le h(\bfY_1|\bfY_2)  + n \nextsc\\
& = h(\bfY_1,\bfY_2)  -h(\bfY_2) + n \nextsc\\
& = h(\tilde\bfX_1, \tilde\bfX_2,\cdots,\tilde\bfX_K, \bfY_1,\bfY_2)- h(\tilde\bfX_1, \tilde\bfX_2,\cdots,\tilde\bfX_K|
\bfY_1,\bfY_2)-h(\bfY_2) + n \nextscnu
\end{align}
where $W_1^K\defn \{W_j\}_{j=1}^K$ and, for each $j$, $\tilde\bfX_j = \bfX_j + \tilde\bfN_j$. Here $\tilde\bfN_j$ is an i.i.d.~sequence and $\tN_j$ is a
Gaussian noise with variance $\sigma_j^2 < \min(1/h_j^2,1/g_j^2)$. Also, $\{\tN_j\}_{j=1}^K$ are mutually independent, and are independent of all other random
variables. Thus,
\begin{align}
n \sum_{i=1}^K R_i
& = h(\tilde\bfX_1, \tilde\bfX_2,\cdots,\tilde\bfX_K, \bfY_1,\bfY_2)
- h(\tilde\bfX_1, \tilde\bfX_2,\cdots,\tilde\bfX_K|
\bfY_1,\bfY_2)-h(\bfY_2) + n \nextscnu \\
& \le h(\tilde\bfX_1, \tilde\bfX_2,\cdots,\tilde\bfX_K, \bfY_1,\bfY_2)
- h(\tilde\bfX_1, \tilde\bfX_2,\cdots,\tilde\bfX_K|
\bfY_1,\bfY_2,\bfX_1,\bfX_2,\cdots,\bfX_K)\nl
&\quad -h(\bfY_2) + n \nextscnu \\
& \le h(\tilde\bfX_1, \tilde\bfX_2,\cdots,\tilde\bfX_K, \bfY_1,\bfY_2)
- h(\tilde\bfN_1, \tilde\bfN_2,\cdots,\tilde\bfN_K|
\bfY_1,\bfY_2,\bfX_1,\bfX_2,\cdots,\bfX_K)\nl
&\quad -h(\bfY_2) + n \nextscnu \\
& \le h(\tilde\bfX_1, \tilde\bfX_2,\cdots,\tilde\bfX_K, \bfY_1,\bfY_2)
 -h(\bfY_2) + n \nextsc \\
& = h(\tilde\bfX_1, \tilde\bfX_2,\cdots,\tilde\bfX_K)
+h(\bfY_1,\bfY_2|\tilde\bfX_1, \tilde\bfX_2,\cdots,\tilde\bfX_K)
-h(\bfY_2) + n \nextscnu \\
\label{eqn:mac_k_y12_given_all_x}
& \le h(\tilde\bfX_1, \tilde\bfX_2,\cdots,\tilde\bfX_K)  -h(\bfY_2) + n \nextsc \\
& = \sum_{j=1}^Kh(\tilde\bfX_j)   -h(\bfY_2) + n \nextsc \\
& \le \sum_{j=2}^Kh(\tilde\bfX_j) + n \nextsc
\label{eqn:mac_k_ub_y_2_by_x_1}
\end{align}
where \eqn{eqn:mac_k_y12_given_all_x} follows similar to \eqn{eqn:gwch_general_ub_for_m_helpers_reconstruction}, and \eqn{eqn:mac_k_ub_y_2_by_x_1} is due to
\begin{equation}
h(\tilde\bfX_1)
\le h( g_1 \bfX_1 + \bfN_2) + n \nextsc
\le h(\bfY_2) + n \nextscnu
\label{eqn:mac_k_ub_for_r_1}
\end{equation}
which is similar to going from \eqn{eqn:gwch_general_ub_for_m_helpers} to \eqn{eqn:gwch_general_ub_for_m_helpers_kill_y2} in Lemma \ref{lemma:gwch_general_ub_for_m_helpers} by using
derivations in \eqn{eqn:gwch_kill_y2_start}-\eqn{eqn:gwch_kill_y2_end}.

On the other hand, for each $j$, we have a bound similar to Lemma~\ref{lemma:gwch_general_ub_for_helper}
\begin{align}
\sum_{i\neq j} H( W_{i})
& =   H( W_{\neq j}) \\
& \le I( W_{\neq j};\bfY_1) + n \nextsc\\
& \le I\left( \sum_{i\neq j} h_i \bfX_i ; \bfY_1\right) + n \nextscnu\\
& =   h\left( \bfY_1\right) - h\left( \bfY_1\Bigg|\sum_{i\neq j} h_i \bfX_i\right) + n \nextscnu\\
& =   h\left( \bfY_1\right) - h\left( h_j \bfX_j + \bfN_1 \right)+ n \nextscnu \\
& \le h( \bfY_1) - h(\tilde\bfX_j) + n \nextsc 
\end{align}
where $W_{\neq j} \defn \{W_i\}_{i=1}^K \backslash \{W_j\}$ which forms the
Markov chain $W_{\neq j} \rightarrow \bfX_{\neq j}\rightarrow \sum_{i\neq j} h_i \bfX_i\rightarrow \bfY_1$.
Therefore, for each $j$, we have 
\begin{align}
h(\tilde\bfX_j) \le h( \bfY_1) -  \sum_{i\neq j} H( W_{i}) + n \nextscnu 
\label{eqn:added-step}
\end{align}

Now, continuing from \eqn{eqn:mac_k_ub_y_2_by_x_1} and incorporating \eqn{eqn:added-step}, we have
\begin{align}
n \sum_{i=1}^K R_i
& \le \sum_{j=2}^Kh(\tilde\bfX_j) + n \nextsc \\
& \le \sum_{j=2}^K \left[ h( \bfY_1) - \sum_{i\neq j} H( W_{i}) \right] + n \nextsc
\end{align}
Noting that $H(W_i)=nR_i$, this is equivalent to,
\begin{align}
n R_1 + (K-1) \sum_{j=1}^K n R_j \le (K-1) h(\bfY_1) + n\nextscnu
\end{align}

We then apply this upper bound for each $i$ by eliminating a different
$h(\tilde\bfX_i)$ each time in the same way that it was done for
$h(\tilde\bfX_1)$ in 
\eqn{eqn:mac_k_ub_for_r_1} and have $K$ upper bounds in total:
\begin{align}
n R_i + (K-1) \sum_{j=1}^K n R_j \le (K-1) h(\bfY_1) + n\nextscnu, 
\quad\quad
i=1,2,\cdots,K
\end{align}
Thus,
\begin{align}
 \Big[K (K-1) +1 \Big]\sum_{j=1}^K n R_j
&  \le K (K-1) h(\bfY_1) + n\nextsc\\
&  \le K(K-1) \left( \frac{n}{2} \log P \right) + n \nextsc
\end{align}
that is,
\begin{align}
D_{s,\Sigma} \le  \frac{ K(K-1)}{K(K-1) +1}
\end{align}
which concludes the converse part of the theorem.

\subsection{Achievable Scheme}

\begin{figure}[t]
\centering
\includegraphics[scale=0.8]{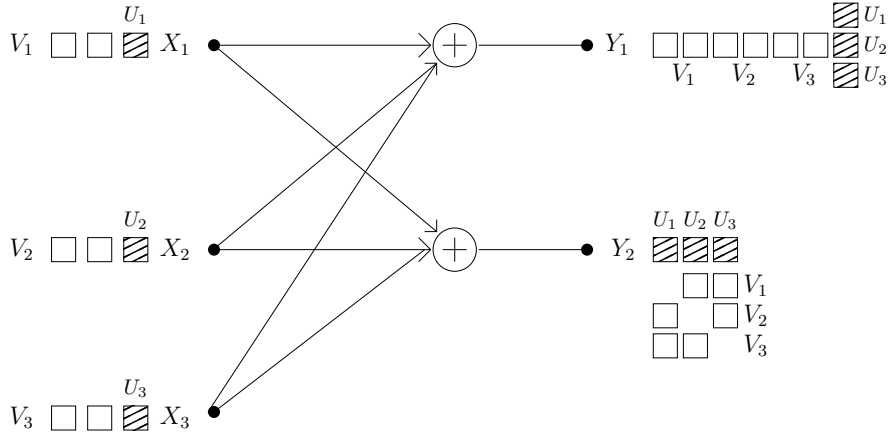}
\caption{Illustration of interference alignment for the $K$-user multiple
access wiretap channel. Here, $K=3$.}
\label{fig:mac_k_ia}
\end{figure}

In the Gaussian wiretap channel with $M$ helpers, our achievability scheme
divided the message signal into $M$ parts, and each one of the $M$ helpers
protected a part at the eavesdropper. On the other hand, in the interference
channel with confidential messages, since each user had its own message to send,
each transmitter sent a combination of a message and a cooperative jamming
signal. We combine these two approaches to propose the following achievability
scheme in this $K$-user multiple access wiretap channel. Each transmitter $i$
divides its message into $(K-1)$ mutually independent sub-signals. In addition,
each transmitter $i$ sends a cooperative jamming signal $U_i$. At the
eavesdropper $Y_2$, each sub-signal indexed by $(i,j)$, where
$j\in\{1,2,\cdots,K\}\backslash \{i\}$, is \emph{aligned} with a cooperative
jamming signal $U_i$. At the legitimate receiver $Y_1$, all of the cooperative
jamming signals are \emph{aligned} in the same dimension to \emph{occupy} as
\emph{small} a signal space as possible. This scheme is illustrated in
Figure~\ref{fig:mac_k_ia} for the case of $K=3$.

We use in total $K^2$ mutually independent random variables which are
\begin{align}
& V_{i,j},\quad i,j\in \{1,2,\cdots,K\}, j\neq i \\
& U_k ,\quad k\in\{1,2,\cdots,K\}
\end{align}
Each of them is uniformly and independently drawn from the same constellation
$C(a,Q)$, where $a$ and $Q$ will be specified later. For each $i\in\{1,2,\cdots,K\}$, we choose the input signal of transmitter $i$ as
\begin{equation}
X_i  = \sum_{j=1,j\neq i}^{K} \frac{g_j}{g_i h_j}  V_{i,j} + \frac{1}{h_i} U_i
\end{equation}
With these input signal selections, observations of the receivers are
\begin{align}
Y_1& = \sum_{i=1}^K \sum_{j=1,j\neq i}^{K} \frac{g_j h_i}{g_i h_j}  V_{i,j}
   + \left[ \sum_{k=1}^K U_k \right] + N_1 \\
Y_2& =\left[ \sum_{i=1}^K \sum_{j=1,j\neq i}^{K} \frac{g_j} { h_j}  V_{i,j} \right]
   + \sum_{j=1}^K \frac{g_j}{h_j} U_j+ N_2 \\
& = \sum_{j=1}^K\frac{g_j}{h_j}\left[U_j +\sum_{i=1,i\neq j}^{K}  V_{i,j}\right]+ N_2
\end{align}

By \cite[Theorem 1]{secrecy_ia5}, we can achieve the following sum secrecy rate
\begin{equation}
\sup \sum_{i=1}^{K} R_i  \ge I(\mathbf{V};Y_1) - I(\mathbf{V};Y_2)
\end{equation}
where $\mathbf{V} \defn \{V_{i,j}:i,j\in\{1,2,\cdots,K\}, j\neq i\}$.

Now, we first bound the probability of decoding error. Note that the
\emph{space} observed at receiver $1$ consists of $(2Q+1)^{K(K-1)}(2 KQ+1)$
points in $K(K-1)+1$ \emph{dimensions}, and the sub-signal in each
\emph{dimension} is drawn from a constellation of $C(a,KQ)$. Here, we use the
property that
$C(a,Q)\subset C(a, K Q)$. By using Khintchine-Groshev theorem of Diophantine
approximation in number theory, we can bound the minimum distance $d_{min}$
between the points in the receiver's \emph{space}, i.e.,  for any $\delta>0$,
there exists a constant $k_\delta$ such that
\begin{equation}
d_{min} \ge \frac{ k_\delta  a}{(KQ)^{K(K-1)+\delta}}
\end{equation}
for almost all rationally independent factors in the $Y_1$ except for a set of Lebesgue measure zero. Then, we can upper bound the probability of decoding error of such a PAM scheme by considering the additive Gaussian noise at receiver $1$ as,
\begin{align}
\pe\left[\bfV\neq\hat{\bfV} \right]
& \le \exp\left(  - \frac{d_{min}^2}{8}\right)  
 \le \exp\left(  -\frac{a^2 k_\delta^2}{8 (KQ)^{2(K(K-1)+\delta)}}\right)
\end{align}
where $\hat{\bfV}$ is the estimate of $\bfV$ by choosing the closest point in the constellation based on observation $Y_1$. For any $\delta>0$, if we choose
$Q = P^{\frac{1-\delta}{2(K(K-1)+1+\delta)}}$ and $a=\gamma P^{\frac{1}{2}}/Q$, where $\gamma$ is a constant independent of $P$, then
\begin{align}
\pe\left[\bfV\neq\hat{\bfV}\right]
& \le \exp\left( -\frac{k_\delta^2 \gamma^2 K^2 P}{8 (KQ)^{2(K(K-1)+\delta)+2}} \right)  
 = \exp\left( -\frac{k_\delta^2 \gamma^2 K^2 P^\delta}{8 K^{2(K(K-1)+\delta)}} \right)
\end{align}
and we can have $\pe\left[\bfV\neq\hat{\bfV}\right]\to 0$  as $P\to\infty$. To satisfy the power constraint at the transmitters, we can simply choose
\begin{equation}
\gamma \le
\min_i
  \frac{1}{
  \sqrt{ \sum_{j=1,j\neq i}^{K} \left(\frac{g_j}{g_i h_j} \right)^2
   + \left(\frac{1}{h_i}\right)^2 }
   }
\end{equation}
By Fano's inequality and the Markov chain $\bfV\rightarrow Y_1\rightarrow\hat{\bfV}$, we know that
\begin{align}
H(\bfV| Y_1)
& \le H(\bfV|\hat{\bfV}) \\
& \le 1 +
\exp\left( -\frac{k_\delta^2 \gamma^2 K^2 P^\delta}{8 K^{2(K(K-1)+1+\delta)}}
\right) \log(2Q+1)^{K(K-1)}
\end{align}
which means that
\begin{align}
I(\bfV;Y_1)
&  = H(\bfV) - H(\bfV|Y_1) \\
&\ge
\left[ 1-
\exp\left( -\frac{k_\delta^2 \gamma^2 K^2 P^\delta}{8 K^{2(K(K-1)+1+\delta)}}
\right) \right]\log(2Q+1)^{K(K-1)} -1
\label{eqn:mac_m_lb_ixy1}
\end{align}
On the other hand,
\begin{align}
  I(\bfV;Y_2)
& \le I\left(\bfV;
\sum_{j=1}^K
      \frac{g_j}{h_j}
        \left[
           U_j +
            \sum_{i=1,i\neq j}^{K}  V_{i,j}
        \right]
\right)\\
& =
H\left(
\sum_{j=1}^K
      \frac{g_j}{h_j}
        \left[
           U_j +
            \sum_{i=1,i\neq j}^{K}  V_{i,j}
        \right]
\right)  -
H\left(
\sum_{j=1}^K
      \frac{g_j}{h_j}
        \left[
           U_j +
            \sum_{i=1,i\neq j}^{K}  V_{i,j}
        \right]
 \Bigg| \bfV
\right)\\
& =
H\left(
\sum_{j=1}^K
      \frac{g_j}{h_j}
        \left[
           U_j +
            \sum_{i=1,i\neq j}^{K}  V_{i,j}
        \right]
\right)  -
H\left(
\sum_{j=1}^K
      \frac{g_j}{h_j}
           U_j
\right)
\label{eqn:mac_k_for_combined_constellation_streams} \\
& \le K \log\frac{ 2 K Q +1 } { 2Q +1 } \\
& \le K \log K
\label{eqn:mac_m_lb_ixy2}
\end{align}
where \eqn{eqn:mac_k_for_combined_constellation_streams} is due to the fact that entropy is maximized by the uniform distribution which takes values over a set
of cardinality $(2KQ +1)^{K}$.

Combining \eqn{eqn:mac_m_lb_ixy1} and \eqn{eqn:mac_m_lb_ixy2}, we obtain
\begin{align}
\sup \sum_{i=1}^K   R_i
&  \ge I(\bfV;Y_1) - I(\bfV;Y_2) \\
&  \ge
\left[ 1-
\exp\left( -\frac{k_\delta^2 \gamma^2 K^2 P^\delta}{8 K^{2(K(K-1)+1+\delta)}}
\right) \right]\log(2Q+1)^{K(K-1)}-1  - K \log K\\
&  ={\frac{K(K-1)(1-\delta)}{K(K-1)+1+\delta}} \left(\frac{1}{2}\log P\right) + o(\log P)
\end{align}
where $o(\cdot)$ is the little-$o$ function. If we choose $\delta$ arbitrarily
small, then we can achieve $\frac{K(K-1)}{K(K-1)+1}$ sum secure \dof
with probability one.

\section{Conclusion}

We determined the secure \dof of several fundamental channel models in one-hop
wireless networks. We first considered the Gaussian wiretap channel with one
helper. While the helper needs to create interference at the eavesdropper, it
should not create too much interference at the legitimate receiver. Our approach
is based on understanding this trade-off that the helper needs to strike. To
that purpose, we developed an upper bound that relates the entropy of the
cooperative jamming signal from the helper and the message rate. In addition, we
developed an achievable scheme based on real interference alignment which aligns
the cooperative jamming signal from the helper in the same \emph{dimension} as
the message signal. This ensures that the information leakage rate is upper
bounded by a constant which does not scale with the power. In addition, to help
the legitimate user decode the message, our achievable scheme renders the
message signal and the cooperative jamming signal distinguishable at the
legitimate receiver. This essentially implies that the message signal can
\emph{occupy} only half of the available space in terms of the degrees of
freedom. Consequently, we showed that the exact secure \dof of the Gaussian
wiretap channel with one helper is $\frac{1}{2}$ by these matching
achieavibility and converse proofs. We then generalized our achievability and
converse techniques to the Gaussian wiretap channel with $M$ helpers, Gaussian
broadcast channel with confidential messages and helpers, two-user Gaussian
interference channel with confidential messages and helpers, and $K$-user
Gaussian multiple access wiretap channel. In the multiple-message settings,
transmitters needed to send a mix of their own messages and cooperative jamming
signals. We determined the exact secure \dof in all of these system models.

\appendix

\section{An Alternative Proof for the Multiplexing Gain of the $K$-User Gaussian Interference Channel}
\label{sec:alternative_proof_of_mg_k_gic}

The original proof for this setting is given by
\cite{multiplexing_gain_of_networks}. Here, we provide an alternative proof for
the
$K=2$ case by using Lemma~\ref{lemma:gwch_general_ub_for_helper}, and then
extend it to the case of general $K$.

For $K=2$, the channel model for the two-user Gaussian interference channel is
\begin{align}
Y_1 & = h_{1,1} X_1 + h_{2,1} X_2 + N_1 \\
Y_2 & = h_{1,2} X_1 + h_{2,2} X_2 + N_2
\end{align}

We start with the definition of the sum rate
\begin{align}
n R_1 + n R_2
&  = H(W_1, W_2) \\
&  = H(W_1, W_2|\bfY_1,\bfY_2)  + I(W_1,W_2; \bfY_1,\bfY_2) \\
&  \le  I(W_1,W_2; \bfY_1,\bfY_2) + n \nextsc \label{eqn:due-to-fano}\\
& =   h(\bfY_1,\bfY_2) - h(\bfY_1,\bfY_2|W_1,W_2)  + n \nextscnu \\
& \le h(\bfY_1,\bfY_2) - h(\bfY_1,\bfY_2|\bfX_1, \bfX_2, W_1,W_2)  + n \nextscnu \\
& \le h(\bfY_1,\bfY_2) + n \nextsc \\
& =   h(\tilde\bfX_1, \tilde\bfX_2,\bfY_1,\bfY_2)
     -h(\tilde\bfX_1, \tilde\bfX_2|\bfY_1,\bfY_2)+ n \nextscnu \\
& \le h(\tilde\bfX_1, \tilde\bfX_2,\bfY_1,\bfY_2)
     -h(\tilde\bfX_1, \tilde\bfX_2|\bfY_1,\bfY_2,\bfX_1,\bfX_2)+ n \nextscnu \\
& \le h(\tilde\bfX_1, \tilde\bfX_2,\bfY_1,\bfY_2)  + n \nextsc \\
& =   h(\tilde\bfX_1, \tilde\bfX_2)
    + h (\bfY_1,\bfY_2| \tilde\bfX_1, \tilde\bfX_2)  + n \nextscnu \\
& \le h(\tilde\bfX_1, \tilde\bfX_2)   + n \nextsc
\end{align}
where the last inequality follows similar to
\eqn{eqn:gwch_general_ub_for_m_helpers_reconstruction} after a derivation
similar to
\eqn{eqn:gwch_reconstruction_start}-\eqn{eqn:gwch_reconstruction_end}, and, for each $j$,
$\tilde\bfX_j = \bfX_j + \tilde\bfN_j$. Here $\tilde\bfN_j$ is an
i.i.d.~sequence of $\tN_j$, which is  Gaussian with variance $\sigma_j^2 <
\min(1/h_{j,1}^2,1/h_{j,2}^2)$. Also, $\{\tN_j\}_{j=1}^K$ are mutually
independent, and are independent of all other random variables.

Then, we apply Lemma~\ref{lemma:gwch_general_ub_for_helper} to characterize the
interference from $X_1$ to transmitter-receiver pair $2$ and from $X_2$ to transmitter-receiver pair $1$
\begin{align}
 n R_1 + n R_2
&  \le  h(\tilde\bfX_1, \tilde\bfX_2) + n \nextscnu \\
&  \le  h(\tilde\bfX_1) + h( \tilde\bfX_2) + n \nextscnu \\
&  \le h(\bfY_2) - H(W_2) + h(\bfY_1)  - H(W_1) + n \nextsc
\end{align}
By noting that $H(W_1)=nR_1$ and $H(W_2)=nR_2$, we have
\begin{align}
2(n R_1 + n R_2)
&  \le h(\bfY_2) + h(\bfY_1)   + n \nextscnu \\
&  \le 2 \left( \frac{n}{2} \log P\right) + n \nextsc
\end{align}
which implies that
\begin{equation}
D_{\Sigma} \defn \lim_{P\to\infty} \sup \frac{ R_1 +  R_2}{\frac{1}{2} \log P} \le 1
\end{equation}
i.e., the multiplexing gain of the two-user Gaussian interference channel is not
greater than $1$.  By the argument in \cite[Proposition
1]{multiplexing_gain_of_networks}, we can conclude that the multiplexing gain of
the $K$-user Gaussian interference channel is at most $\frac{K}{2}$.


\end{document}